\documentclass[preprint,12pt]{elsarticle}

\usepackage{amssymb}
\usepackage{amsmath}

\usepackage{graphicx} 
\usepackage{float}
\usepackage[margin=1in]{geometry}
\usepackage[nottoc]{tocbibind}
\usepackage[nolist]{acronym}
\usepackage{amsmath}
\usepackage{natbib}
\usepackage{wrapfig}
\usepackage{float}
\usepackage{caption}
\usepackage{subcaption}
\usepackage[section]{placeins}
\usepackage{xcolor}
\usepackage{tcolorbox}
\usepackage{algorithm}
\usepackage{algorithmic}

\usepackage{amsmath}
\usepackage{amssymb}
\usepackage{graphicx}
\usepackage{cite}
\usepackage{hyperref}
\usepackage{enumitem}
\usepackage{fixltx2e}
\usepackage{todonotes}
\usepackage{pdflscape} 

\usepackage{booktabs,multirow}
\usepackage{adjustbox} 

\begin{acronym}
    \acro{NO}{Neural Operator}
    \acro{FNO}{Fourier Neural Operator}
    \acro{ML}{Machine Learning}
    \acro{PDE}{Partial Differential Equation}
    \acro{PINNs}{physics-informed neural networks}
    \acro{DeepONets}{Deep Operator Networks}
    \acro{DOF}{Degree of Freedom}
\end{acronym}

\journal{Computer Methods in Applied Mechanics and Engineering}

\begin{document}

\begin{frontmatter}



\title{Fourier Neural Operators for Structural Dynamics Models: Challenges, Limitations and Advantages of Using a Spectrogram Loss}

\author[inst1]{Rad Haghi\corref{cor1}}
\cortext[cor1]{Corresponding author:}
\ead{rad.haghi@tufts.edu}
\author[inst1]{Bipin Gaikwad}
\author[inst1,inst2]{Abani Patra}

\affiliation[inst1]{organization={Tufts Institute for AI},
            city={Medford},
            postcode={02155}, 
            state={MA},
            country={USA}}
\affiliation[inst2]{organization={Dept. of Computer Science, Tufts University},
            city={Medford},
            postcode={02155}, 
            state={MA},
            country={USA}}

\begin{abstract}
Fourier Neural Operators (FNOs) have emerged as promising surrogates for partial differential equation solvers. In this work, we extensively tested FNOs on a variety of systems with non-linear and non-stationary properties, using a wide range of forcing functions to isolate failure mechanisms.
FNOs stand out in modeling linear systems, regardless of complexity, while achieving near-perfect energy preservation and accurate spectral representation for linear dynamics. However, they fail on non-linear systems, where the failure manifests as artificial energy dissipation and manipulated frequency content. This limitation persists regardless of training dataset size, and we discuss the root cause through discretization error analysis. Comparison with LSTM as the baseline shows FNOs are superior for both linear and non-linear systems, independent of the training dataset size.
We develop a spectrogram-based loss function that combines time-domain Mean Squared Error (MSE) with frequency-domain magnitude and phase errors, addressing the low-frequency bias of FNOs. This frequency-aware training eliminates artificial dissipation in linear systems and enhances the energy ratios of non-linear systems. 
The IEA 15MW turbine model validates our findings. Despite hundreds of degrees of freedom, FNO predictions remain accurate because the turbine behaves in a predominantly linear regime. Our findings establish that system non-linearity, rather than dimensionality or complexity, determines the success of FNO. These results provide clear guidelines for practitioners and challenge assumptions about FNOs' universality.
\end{abstract}







\begin{keyword}



Fourier Neural Operator \sep Structural Dynamics \sep Complex Linear Systems \sep Spectrogram Loss

\end{keyword}

\end{frontmatter}

\section{Introduction}

Neural operators have recently emerged as a powerful framework 
with a formulation designed to learn mappings between infinite-dimensional function spaces. By potentially enabling mesh-independent and resolution-invariant approximations of partial differential equation (PDE) solution operators, they promise major advances in scientific computing. Among these methods, the Fourier Neural Operator (FNO) \citep{li2021fourier} has drawn particular attention, with reported computational speedups of up to three orders of magnitude compared to conventional solvers. FNOs 
claim resolution-independent solutions by combining the compactness of Fourier space approximations and the use of convolution theorem.
By parameterizing integral operators in Fourier space 
and using convolutions FNOs theoretically achieve resolution-independent representations with quasilinear $\mathcal{O}(N \log N)$ complexity \citep{li2021fourier,kovachki2021neural}.
Despite this promise, FNOs have seen little systematic evaluation in structural dynamics, where time-varying non-linear mechanical systems are the norm. Applications such as modal analysis, vibration response, and damage detection remain largely unstudied by FNOs.  Structural systems frequently undergo parameter changes due to degradation, fatigue, or damage accumulation. These systems also display non-linearities from large deflections, contact mechanics, or constitutive effects. The combination of non-linearity and temporal variation represents a critical test case for neural operators and a very promising area of application if successful.

Recent research has revealed important limitations to FNOs. For example, \citet{lu2022comprehensive} showed that FNOs are highly sensitive to input perturbations, with errors increasing by four orders of magnitude under only 0.1\% noise contamination. They also show that the success in flow modeling is restricted to smooth laminar flows. Spectral analysis by \citet{oommenIntegratingNeuralOperators2025} confirmed a strong low-frequency bias of FNOs, and traced it to the truncation of high-frequency modes and the resulting poor resolution of sharp gradients. In other words, the global Fourier basis assumes quasi-stationarity, which is incompatible with systems demonstrating strong non-linearity or changing parameters.
%
\citet{kovachkiNeuralOperatorLearning2023} demonstrated the universal approximation properties of the FNOs and illustrated use in many physics models; however, several of the claims have since been challenged. \citet{bartolucci2024discretization} showed that discretization invariance often fails in practice, with models trained at coarse resolution performing inconsistently when evaluated at finer grids. \citet{lanthaler2024data} established that as the operators become more complex, the sample requirements grow exponentially, which sets limits on the practical applicability of FNOs.

%
Efforts to extend FNO capabilities have produced mixed outcomes. Multigrid tensorized FNOs \citep{kossaifi2023multigrid} achieve up to 150$\times$ compression with improved performance on Navier–Stokes flows, and spectral boosting strategies \citep{qinBetterUnderstandingFourier2024} reduce errors by up to 71\% by partially recovering high-frequency content. These improvements remain limited to fluid applications and do not directly address the challenges posed by non-linear, time-varying mechanical systems.

In this work, we systematically investigate the successes and limitations of FNOs in non-linear mechanical systems with evolving parameters. Using controlled experiments on two-degree-of-freedom (2DOF) mass–spring models, we isolate the effects of non-linearity and temporal variation. We examine four configurations—linear springs, linear springs with softening, non-linear springs, and non-linear springs with softening—across dataset sizes spanning three orders of magnitude. This design allows us to separate the effects of system complexity, spectral characteristics, and data availability on FNOs' accuracy.

Our results show a clear separation. FNOs achieve high accuracy for linear systems, even with parameter variation, but they fail catastrophically for non-linear systems, regardless of training data size. This breakdown is architectural rather than due to lack of data or training strategy. Thus, FNOs successfully model complex linear systems, including the IEA 15MW reference wind turbine; however, they cannot capture representative 2DOF non-linear dynamical system. It means, non-linearity or non-stationarity, not system size or complexity, is the decisive factor.

We also explore the use of    frequency-aware losses to overcome   the low-frequency bias \citet{oommenIntegratingNeuralOperators2025}. Including spectrogram losses provides differential weighting of high-frequency modes reducing this bias, but remains insufficient to yield accurate predictions for non-linear systems. These findings indicate that the loss refinement alone may be inadequate, and fundamental architectural changes in FNO are required for  non-linear systems.

The contributions of this paper are fourfold. (i) We provide a systematic characterization of FNO performance in mechanical systems, identifying quantitative thresholds for applicability. (ii) We show that non-linearity, rather than complexity or dimensionality, is the core limitation of current FNO architectures. (iii) We link dataset size to accuracy, demonstrating poor data efficiency in non-linear systems. (iv) We offer practical guidelines for when FNOs can be expected to succeed and when they would fail.

The remainder of this paper is organized as follows. Section~\ref{sec:meth_1} reviews the theoretical background of neural operators and FNOs, system models, {\it a priori } error estimates, and spectrogram loss. Section \ref{sec:meth_2} presents the FNO architecture, the reference case, and evaluation metrics. Section~\ref{sec:data} describes our experimental setup and data generation. Section~\ref{sec:result} reports and discusses results across system configurations, dataset sizes, and loss functions. Section~\ref{sec:conc} concludes with future directions.

\FloatBarrier

\section{FNOs and Modeling Structural Dynamics}
\label{sec:meth_1}

We design a comprehensive framework to map the performance boundaries of FNOs and to isolate specific failure modes. The framework integrates controlled synthetic experiments and baseline comparisons, providing a picture of FNO capabilities and limitations.

The experimental design contains six interconnected components:

\textbf{Component 1: Controlled data generation and system design.}
We generate data from two-degree-of-freedom (2DOF) mass–spring systems with controlled parameters. Four configurations motivated by engieering use cases \ref{tab:applications} vary the source of complexity: (i) linear with constant stiffness; (ii) linear with exponentially softening stiffness to mimic material degradation; (iii) non-linear with cubic stiffness (geometric non-linearity); and (iv) non-linear with time-varying parameters (combined complexity). Each configuration is exercised under three forcing regimes—low-frequency (0.05–0.1\,Hz), high-frequency (1–2\,Hz), and broadband (0.04–2.1\,Hz)—resulting in twelve test conditions. This factorial design separates the effects of non-linearity and temporal variation and examines their interaction.

\textbf{Component 2: Multi-level training and data-efficiency analysis.} 
We train on nested datasets of increasing size (64, 128, 256, 512, 1024, 2048 samples), where each larger set is a superset of the smaller ones, enabling consistent comparison. This helps us assess how performance scales with training data size and identifies the minimum data needed for reliable generalization. To isolate the effect of training dataset size variation from that of architecture, the FNO hyperparameters are held fixed across experiments. Training uses early stopping with learning-rate scheduling to ensure convergence while limiting overfitting.

\textbf{Component 3: Comprehensive failure-mode testing.}
Beyond standard accuracy, we examine three out-of-distribution input forces: (i) frequency-sweep (chirp) responses from 0.01–2\,Hz to identify critical frequency boundaries; (ii) impact responses approximating impulse loads to assess transient behavior; and (iii) sudden load changes to test discontinuities. These tests expose failure mechanisms that pointwise metrics can miss.

\textbf{Component 4: Baseline model.}
We implement an LSTM baseline to compare FNO performance with a non-spectral sequential model, using identical datasets and splits. Details of the LSTM appear in Section~\ref{sec:lstm}.

\textbf{Component 5: Frequency-aware losses and mitigation strategies.}
FNOs have a well-documented low-frequency bias \citet{oommenIntegratingNeuralOperators2025}. Therefore, we evaluate a spectrogram-based loss combining time-domain MSE with frequency-domain magnitude and phase terms (Section~\ref{sec:spectroloss}) to bypass this bias. This tests whether training modifications can partially mitigate architectural bias.

\textbf{Component 6: Validation on a complex linear system.}
To separate non-linearity from dimensional complexity, we validate on the IEA 15\, MW reference wind turbine (hundreds of DOFs; predominantly linear). The dataset includes six SCADA channels and structural responses under Design Load Case (DLC)~1.2 (normal operation) and DLC~6.4 (parked/idling). This comparison decouples system complexity from the non-linearity limitation.
Across all conditions, we report: (i) an energy ratio in the signals, (ii) a PSD-shape error via spectral NRMSE (frequency-distribution accuracy), and (iii) a Spectral Coherence error to combine frequency and phase accuracy in the Fourier domain. Section~\ref{sec:meth_error_metric} formalizes the metrics. In the following, we describe the operators, losses, and periori errors.

\subsection{Theoretical Foundations of Fourier Neural Operators}
\label{sec:fno_theory}
FNOs learn approximate maps between infinite-dimensional function spaces via compositions of integral operators and non-linear activations. Formally, let $\mathcal{A}$ and $\mathcal{U}$ be Banach spaces of functions on a bounded domain $D \subset \mathbb{R}^d$. A neural operator $\mathcal{G}_\theta:\mathcal{A}\!\to\!\mathcal{U}$ has the structure \citep{li2021fourier,kovachkiNeuralOperatorLearning2023}
\begin{equation}
\mathcal{G}_\theta = Q \circ \sigma_T\!\big(W_{T-1} + \mathcal{K}_{T-1} + b_{T-1}\big) \circ \cdots \circ \sigma_1\!\big(W_0 + \mathcal{K}_0 + b_0\big) \circ P,
\label{eq:G}
\end{equation}
where $P:\mathbb{R}^{d_a}\!\to\!\mathbb{R}^{d_{v_0}}$ lifts inputs, $Q:\mathbb{R}^{d_{v_T}}\!\to\!\mathbb{R}^{d_u}$ projects to outputs, and $\mathcal{K}_t$ are kernel integral operators, $W_t$ are the local linear operators, $b_t$ are the bias functions, and $t$ is the layer in the model architecture.

For translation-invariant kernels $\kappa(x,y)=\kappa(x-y)$, convolution $u(x)=\int_D \kappa(x-y)v(y)\,dy$ reduces to multiplication in the Fourier domain:
\begin{equation}
u = \mathcal{F}^{-1}\!\left(\mathcal{F}[\kappa]\cdot \mathcal{F}[v]\right).
\label{eq:f_f_inv}
\end{equation}
FNO parameterizes this operation directly in Fourier space with learnable complex weights $R_\theta\in\mathbb{C}^{k_{\max}\times m \times n}$, truncating to $k_{\max}$ modes for tractability \citep{li2021fourier}. The typical layer update is:

\begin{equation}
v_{l+1}(x)=\sigma\!\left(W_l v_l(x)+\mathcal{F}^{-1}\!\big(R_\theta \cdot \mathcal{F}(v_l)\big)(x)+b_l(x)\right).
\label{eq:v_sigma}
\end{equation}

\noindent with Fast Fourier Transform (FFT)-based implementations, the per-layer complexity scales as $\mathcal{O}(N\log N + k_{\max} d_v N)$, and the frequency-domain parameterization supports 
generalization to grids unseen during training. 

FNOs can be viewed as learnable spectral methods: Fourier layers realize global convolutions similar to spectral Galerkin schemes; when $R_\theta$ is identity and mode truncations match classical bases, the model reduces to a traditional spectral method \citep{brandstetter2022spectral}. Pseudospectral connections arise between modes when commuting between spectral and spatial domains via FFT, with non-linearities introducing aliasing in physical space. Recent work also adapts classical de-aliasing strategies, including the 3/2 rule and spectral normalization \citep{mccabeStabilityAutoregressiveNeural2023}. 



Time-domain losses (e.g., MSE) can under-penalize spectral distortions that are critical in vibration and dynamics \citep{worden2007nonlinear}. To address the FNO low-frequency bias \citep{oommenIntegratingNeuralOperators2025}—and to supervise time-varying content—we adopt a spectrogram-based loss that couples magnitude and phase.
 
\subsection{Spectrogram Loss for Frequency-Aware Learning}
\label{sec:spectroloss}
For a discrete signal $x[n]$, the short-time Fourier transform (STFT) is
\begin{equation}
X[m,k] = \sum_{n=0}^{N-1} x[n]\, w[n-mH]\, e^{-j 2\pi k n/N},
\end{equation}
with frame index $m$, frequency bin $k$, window $w[\cdot]$ (Hann), hop $H$, and FFT size $N$ \citep{allenShortTermSpectral1977}. The total spectral loss is
\begin{equation}
\mathcal{L}_{\text{spectral}} = \lambda_{\text{mag}}\,\mathcal{L}_{\text{mag}} + \lambda_{\text{phase}}\,\mathcal{L}_{\text{phase}},
\end{equation}
\noindent where the magnitude term estimates spectral convergence \citep{yamamoto2020parallel}:
\begin{equation}
\mathcal{L}_{\text{mag}} = \frac{\big\|\,|\hat{X}| - |X|\,\big\|_F}{\big\|\,|X|\,\big\|_F + \epsilon},
\end{equation}
\noindent and the phase term penalizes circular phase error:
\begin{equation}
\mathcal{L}_{\text{phase}} = \frac{1}{MK}\sum_{m,k}\!\left[\operatorname{angle}\!\big(e^{j(\phi_{\hat{X}}[m,k]-\phi_X[m,k])}\big)\right]^2.
\end{equation}

\noindent where $\phi_{{X}}$ is the phase out of the STFT process.  

We restrict supervision to physically relevant frequencies with a mask up to $f_{\max}=4.0$\,Hz, which covers the dominant 2DOF dynamics while excluding high-frequency noise. Unless otherwise stated, we use $N\!=\!1024$ (frequency resolution $\approx 0.024$\,Hz), $H\!=\!512$ (50\% overlap), and $\lambda_{\text{mag}}\!:\!\lambda_{\text{phase}}=1.0\!:\!0.1$. Signals are de-normalized to physical units before STFT. The total loss combines time and frequency domains:
\begin{equation}
\mathcal{L}_{\text{total}} = \alpha\,\mathcal{L}_{\text{MSE}} + (1-\alpha)\,\mathcal{L}_{\text{spectral}},
\label{eq:total_loss}
\end{equation}
with $\alpha=0.8$ chosen through empirical tests. The STFT provides supervision at intermediate time–frequency scales, complementing global (pure Fourier) and local (time-domain) objectives.

\subsection{{\it a priori} Error Analysis}
\label{sec:piriori_error}

FNO approximation error arises from two fundamental sources that interact in complex ways for non-linear systems. Recent theoretical work provides rigorous bounds on both components.

The first source is the mode truncation error. FNOs approximate operators by retaining only a finite number of Fourier modes up to some maximum frequency $k_{max}$, discarding all higher-frequency information. Kovachki et al. \citep{kovachkiUniversalApproximationError2021} prove that this truncation creates a fundamental error floor: the FNO approximation error cannot be smaller than the norm of the discarded high-frequency modes. This bound has important practical implications. When the true output contains mostly low-frequency content with rapidly decaying Fourier coefficients, the truncation error is small and decreases exponentially as more modes are retained. However, when the output contains substantial high-frequency content, this error decreases much more slowly—polynomially rather than exponentially—with the number of retained modes. As a result, even with many Fourier modes, FNOs cannot accurately represent outputs with significant high-frequency components. 

\citet{kovachkiUniversalApproximationError2021} distinguishes between two levels of approximation in FNOs. At the theoretical level, FNOs provide a spectral approximation of continuous operators through the universal approximation theorem. At the practical level, however, FNOs must be implemented using discrete Fourier transforms with finite mode truncation. It is this second, practical limitation—the necessity of truncating Fourier modes—that creates the fundamental error floor discussed above. While the theoretical FNO with infinite modes could approximate any continuous operator, the practical implementation with finite $k_{max}$ faces hard limits when the target operator generates high-frequency content.

The second source is the discretization and aliasing error in the training data used in practical implementations. \citet{lanthalerDiscretizationErrorFourier2024a} recently established rigorous bounds on this aliasing error. Their main conclusion is as follows: the discretization error in the $\ell^2$ norm scales as $\mathcal{O}(N^{-s})$. Here, $N$ is the grid resolution and $s$ is the Sobolev regularity of the input functions \citep{lanthalerDiscretizationErrorFourier2024a}.

This theoretical framework explains our experimental results with the 2DOF non-linear system. The discretization error bound is:

\begin{equation}
\frac{1}{N^{d/2}}\|v_t - v_t^N\|_{\ell^2(n \in [N]^d)} \leq CN^{-s}
\end{equation}

\noindent Here $v_t$ is the true Fourier state at layer $t$. The term $v_t^N$ is its discretized approximation. The constant $C$ depends on network parameters, activation smoothness, and input properties. The main takeaway is that the convergence rate depends largely   on the regularity $s$ as $N$ gets large.

\FloatBarrier

\section{Methodology}
\label{sec:meth_2}
Having established the theoretical foundations of FNOs, spectrogram loss, and {\it a priori} error analysis, we now turn to implementation details. This section presents the specific FNO architecture employed in our experiments, introduces the LSTM baseline model for comparison, and defines the error metrics used for a posteriori performance evaluation.

\subsection{FNO Architecture}

Our surrogate is a 1D Tensorized Fourier Neural Operator (TFNO1d) with four spectral blocks. Tensorization (Tucker factorization of weight tensors) reduces parameters while preserving expressiveness; in our tests, an untensorized FNO yields comparable accuracy but trains slower.

\begin{figure}
    \centering
    \includegraphics[width=0.7\linewidth]{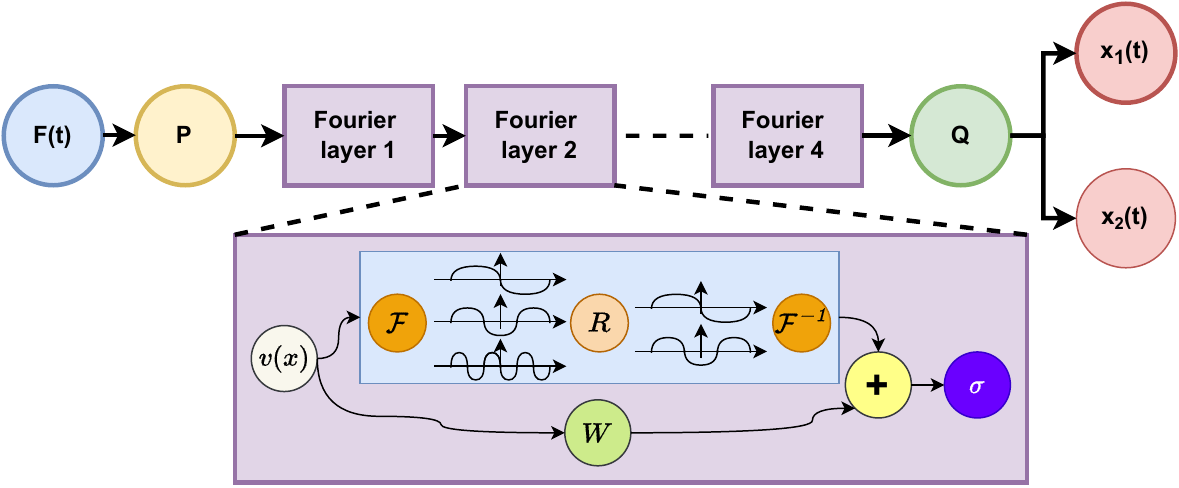}
    \caption{The FNO architecture employed for the 2DOF systems. The input $F(t)$ is the forcing function, and the outputs are displacements ($x_1$ and $x_2$). We used a 4-layer FNO architecture with 1024 modes for each layer. The drawing shows schematically the process that is explained in Section \ref{sec:fno_theory} }
    \label{fig:FNO_schematic}
\end{figure}

A lifting layer maps the inputs—either a single-channel forcing (2DOF case) or seven channels (OWT case)—to a latent width of 128. Each TFNO block applies: (i) a spectral convolution that transforms features to Fourier space, truncates to the lowest 1024 modes along the temporal dimension, applies learned linear transforms, and inverts the transform; and (ii) a channel MLP (pointwise $1\!\times\!1$ convolution) for local mixing. The spectral and local branches are merged through residual connections. Pre-activation is used (non-linearities and any normalization before certain linear maps) for training stability. A dropout with a rate of 0.2 is applied in the channel-MLP to reduce overfitting. After four blocks, we remove any domain padding and project from 128 hidden channels to two output channels. This design couples nonlocal spectral modeling with local refinements, balancing accuracy and efficiency. Figure \ref{fig:FNO_schematic} provides a schematic drawing of the FNO architecture used in this study for the 2DOF system. In Figure \ref{fig:FNO_schematic}, we also show the process that we explained in Sectoin \ref{sec:fno_theory} and Equations \eqref{eq:G} to \eqref{eq:v_sigma}.

\subsection{Long Short-Term Memory Network as Baseline Architecture}
\label{sec:lstm}
The evaluation of Fourier Neural Operators (FNOs) for dynamical system modeling requires comparison against established sequence learning methods. A Long Short-Term Memory (LSTM) network \citep{hochreiter1997lstm} is adopted as the primary baseline due to its proven capability in capturing complex temporal dependencies in non-linear systems. LSTMs enhance traditional recurrent neural networks through input, forget, and output gates that regulate information flow, enabling the learning of both short- and long-term dynamics while mitigating vanishing gradient issues.

LSTMs possess a theoretical capacity to approximate non-linear dynamical systems, as established by the universal approximation theorem for recurrent networks \citep{schafer2006recurrent}. In this study, the LSTM learns the mapping 
$F: L([0,T]) \rightarrow L^2([0,T])$
from the forcing input to the displacement responses of the two masses. In contrast to FNOs—which learn global operators through spectral convolutions in the frequency domain—LSTMs operate in the time domain via recurrent updates, providing a robust baseline for evaluating the advantages of operator learning in dynamic response prediction. A two-layer LSTM is followed by two fully connected layers to model temporal dynamics and perform non-linear regression from the latent features to displacement outputs.

\subsection{Error Metrics}
\label{sec:meth_error_metric}
We introduce here measures of error   appropriate for the structural dynamics systems of interest here.
We evaluate both time-domain energy preservation and frequency-domain components error to investigate the FNO output comprehensively. 

\subsubsection{Root-mean-square (RMS) energy ratio}
For predicted displacement $\mathbf{u}_{\text{pred}}(t)\!\in\!\mathbb{R}^{n\times T}$ and ground truth $\mathbf{u}_{\text{true}}(t)$,
\begin{align}
\mathrm{RMS}_i^{\text{pred}} &= \sqrt{\frac{1}{T}\sum_{t=1}^{T} u_{i,\text{pred}}^2(t)},\quad
\mathrm{RMS}_i^{\text{true}} = \sqrt{\frac{1}{T}\sum_{t=1}^{T} u_{i,\text{true}}^2(t)}, \quad i=1,\dots,n,\\
\mathcal{R}_i &= \frac{\mathrm{RMS}_i^{\text{pred}}}{\mathrm{RMS}_i^{\text{true}}}.
\end{align}
Here $\mathcal{R}_i\!=\!1$ indicates perfect energy preservation; $\mathcal{R}_i<1$ implies dissipation; $\mathcal{R}_i>1$ implies spurious energy.

\subsubsection{PSD-shape error via NRMSE}
We estimate PSD via Welch’s method \citep{welch1967,stoica2005spectral}. For a series $x[n]$ of length $N$, with $K$ overlapping segments of length $L$ and overlap ratio $\alpha$,
\begin{equation}
x_k[m] = x[m + k(1-\alpha)L], \quad m=0,\dots,L-1,
\end{equation}
apply a window $w[m]$ (Hann) and compute
\begin{equation}
X_k(f) = \sum_{m=0}^{L-1} x_k[m]\, w[m]\, e^{-j2\pi f m/f_s},
\end{equation}
then average
\begin{equation}
\hat{S}_x(f) = \frac{1}{K W}\sum_{k=0}^{K-1} |X_k(f)|^2,\qquad W=\sum_{m=0}^{L-1} w^2[m].
\end{equation}
The PSD-shape error is
\begin{equation}
\mathrm{NRMSE}_{\text{PSD}} = \frac{\sqrt{\frac{1}{N_f}\sum_{j=1}^{N_f}\!\left[\hat{S}_{\text{pred}}(f_j)-\hat{S}_{\text{true}}(f_j)\right]^2}}{\sqrt{\frac{1}{N_f}\sum_{j=1}^{N_f}\!\hat{S}_{\text{true}}^2(f_j)}} \times 100\%.
\end{equation}
We report the mean across test samples. Unless noted, we use $L=\min(256,T/4)$ with 50\% overlap and normalize frequency by the Nyquist rate.

\subsubsection{Spectral coherence}
To capture frequency-specific linear correlation, we compute magnitude-squared coherence \citep{virtanen2020scipy}:
\begin{equation}
C_{\hat{x}x}(f) = \frac{|G_{\hat{x}x}(f)|^2}{G_{\hat{x}\hat{x}}(f)\, G_{xx}(f)},
\end{equation}
where $G_{\hat{x}x}$ is the cross-spectrum and $G_{\hat{x}\hat{x}}$, $G_{xx}$ are autospectra, all estimated with Welch’s method \citep{welch1967}. Coherence complements amplitude-based metrics (e.g., energy ratio) by isolating frequency bands where phase-locked behavior is captured or missed.

\FloatBarrier

\section{Data Generation and Experimental Design}
\label{sec:data}

We utilized two datasets to evaluate FNO performance boundaries. First, we generate synthetic data from controlled 2DOF mass–spring–damper systems in which all parameters are known and systematically varied to isolate FNOs' failure modes. Second, we incorporate simulated data from the IEA 15\,MW reference wind turbine to assess FNO performance on a complex but predominantly linear system.

\subsection{Two-Degree-of-Freedom System Formulation}

\begin{figure}
    \centering
    \includegraphics[width=0.5\linewidth]{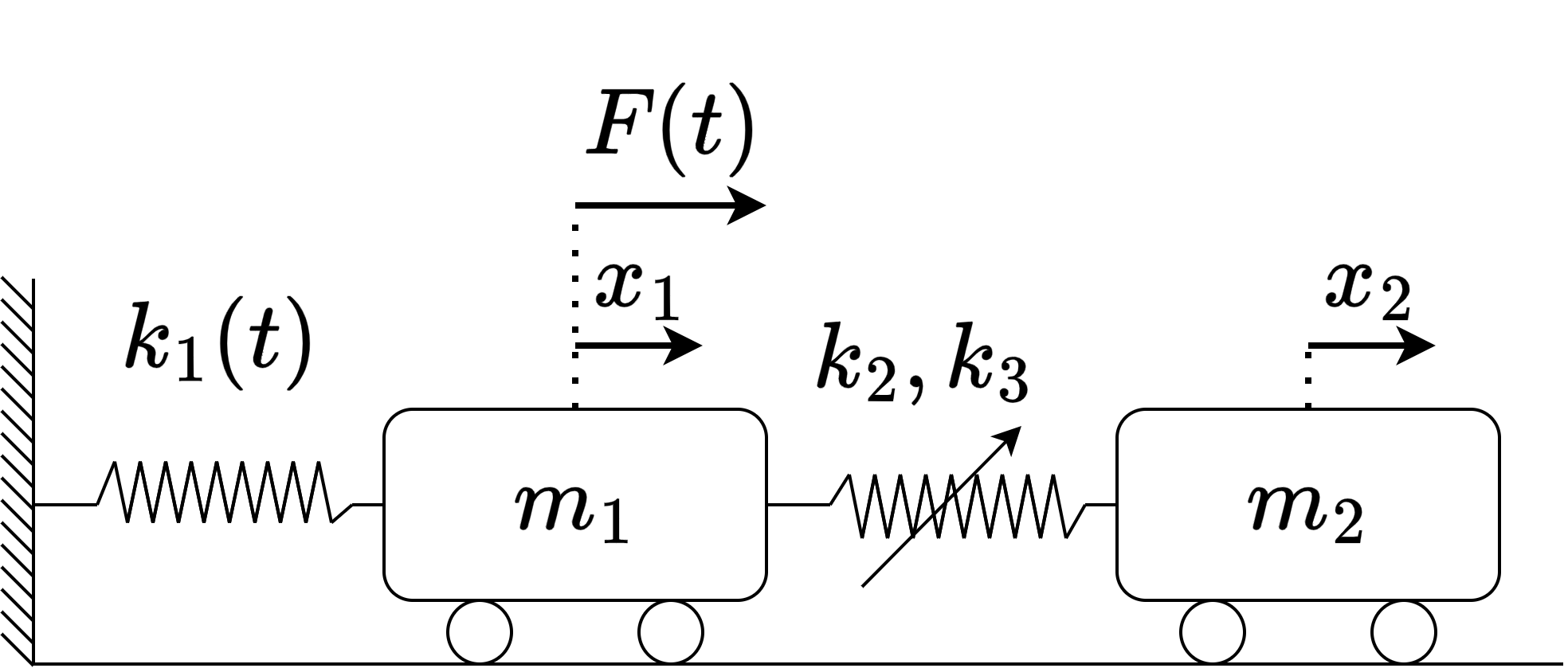}
    \caption{Schematic of the 2DOF mass–spring–damper system. $k_1(t)$ can be set to be a function of time (softening), $k_2$ is a linear spring, and $k_3$ is the cubic non-linearity factor.}
    \label{fig:2DOF}
\end{figure}

The 2DOF system serves as the primary testbed because it can exhibit both linear and non-linear dynamics while remaining analytically tractable (Figure~\ref{fig:2DOF}). The system comprises two masses, $m_1=5$\,kg and $m_2=10$\,kg, connected by springs whose stiffnesses may be time-varying and/or non-linear. The $1{:}2$ mass ratio reflects a common engineering regime—avoiding the trivial equal-mass case and extreme ratios that induce numerical stiffness.

The equations of motion are
\begin{align}
m_1\ddot{x}_1 + k_1(t)\,x_1 + k_2(x_1 - x_2) + k_3(x_1 - x_2)^3 &= F(t), \\
m_2\ddot{x}_2 + k_2(x_2 - x_1) + k_3(x_2 - x_1)^3 &= 0,
\end{align}
where $k_1(t)$ is the wall-spring stiffness (possibly time-varying), $k_2=50$\,N/m is the coupling stiffness, $k_3=20$\,N/m$^{3}$ introduces a cubic non-linearity when present, and $F(t)$ acts on the first mass.

We study four configurations that separate the roles of non-linearity and temporal variation:

\begin{table}[H]
\centering
\caption{2DOF system configurations for systematic complexity isolation}
\begin{tabular}{clll}
\hline
\textbf{Case} & \textbf{Configuration} & \textbf{Stiffness Model} & \textbf{Physical Representation} \\
\hline
1 & Linear & $k_1 = 30$ N/m (constant) & Ideal elastic system \\
  & & $k_3 = 0$ & \\
2 & Linear with & $k_1(t) = 10 + 30e^{-0.05t}$ N/m & Material fatigue/aging \\
  & Softening & $k_3 = 0$ & without non-linearity \\
3 & non-linear & $k_1 = 30$ N/m (constant) & Large deflections or \\
  & & $k_3 = 20$ N/m³ & contact non-linearity \\
4 & non-linear with & $k_1(t) = 10 + 30e^{-0.05t}$ N/m & Real damaged structures \\
  & Softening & $k_3 = 20$ N/m³ & with geometric effects \\
\hline
\end{tabular}
\label{tab:cases}
\end{table}

For the softening cases, $k_1(t)$ decays from $40$\, N/m to $\approx 10$\, N/m by $t=100$\,s, a $\,\sim\!74\%$ reduction, so the degradation is largely complete within the first half of the $200$\,s simulation window. We test three forcing regimes, to examine the spectral behavior of the system:

\begin{table}[H]
\centering
\caption{Forcing frequency configurations for spectral analysis}
\begin{tabular}{clll}
\hline
\textbf{Config} & \textbf{Frequencies} & \textbf{Regime} & \textbf{Testing Purpose} \\
\hline
A & $f_1 \sim \mathcal{U}(0.04, 1.1)$ Hz & Broadband & Generalization across \\
  & $f_2 \sim \mathcal{U}(0.09, 2.1)$ Hz & & frequency spectrum \\
B & $f_1 = 1$ Hz, $f_2 = 2$ Hz & High-frequency & Near-resonance dynamics \\
  & & & and mode coupling \\
C & $f_1 = 0.05$ Hz, $f_2 = 0.1$ Hz & Low-frequency & Long-term coherence \\
  & & & and phase tracking \\
\hline
\end{tabular}
\label{tab:freq_conf}
\end{table}

\noindent where each sample uses the excitation:

\begin{equation}
F(t) = 20 A_{\sin}\sin(2\pi f_1 t + \phi_1) + 16 A_{\cos}\cos(2\pi f_2 t + \phi_2),
\label{eq:force}
\end{equation}
with $A_{\sin},A_{\cos}\sim\mathcal{U}(0,1)$ and $\phi_1,\phi_2\sim\mathcal{U}(0,2\pi)$. Combined with the four system configurations, this yields $12$ experimental conditions.

We generate datasets of size $2^n$ for $n\in\{6,7,8,9,10,11,12\}$ (i.e., 64–4096 samples). We first generate 4096 samples and obtain smaller sets as deterministic subsets, while employing a fixed seed number. Models are trained on 80\% random subsets of each training size. For the final evaluation, all models are tested on the same held-out data. It means the samples 2049–4096 from the 4096-sample dataset (never used in training), enabling fair data-efficiency comparisons.

Each sample is a 200\,s trajectory sampled at 25\, Hz (5000 points). Initial conditions are $x_1(0)=0.1$\,m, $\dot{x}_1(0)=0$\,m/s, $x_2(0)=0.2$\,m, $\dot{x}_2(0)=0$\,m/s to ensure non-trivial free response. We chose the sampling rate to Nyquist for all tested frequencies. We utilized adaptive Runge–Kutta (RK45, SciPy) for time integration; therefore, the numerical error is negligible relative to model error.

The four 2DOF system configurations model diverse engineering applications, as summarized in Table \ref{tab:applications}.

\begin{table}
\centering
\caption{Real-world applications inspiring 2DOF system configurations}
\label{tab:applications}
\small
\begin{tabular}{p{3cm}p{7cm}p{4cm}}
\hline
\textbf{Configuration} & \textbf{Engineering Applications} & \textbf{Key References} \\
\hline
Case 1 & 
- Quarter-car suspensions   \newline
- Tuned mass dampers   &
\citep{muzakkirahamedStructuralDynamicAnalysis2022} \newline
\citep{febbo2008optimization} \\
\hline
Case 2& 
- Reinforced concrete  \newline
- Rubber bushings &
\citep{jiaImproved2DOFModel2023} \newline
\citep{gjbush2023aging} \\
\hline

Case 3 & 
- Progressive vehicle springs \newline
- Cable-stayed bridges &
\citep{wang2008parametric} \newline
\citep{soto2006vibration}\\
\hline

Case 4 & 
- Seismic base isolators  \newline
- Mining equipment suspensions  & 
\citep{saotome2019critical} \newline
\citep{blood2018evaluation} \\
\hline

\end{tabular}
\end{table}

These applications demonstrate that each configuration captures distinct physical phenomena: material degradation (softening), geometric effects (non-linearity), or their combination. The selection of an appropriate model depends on the dominant physics rather than system complexity.

\subsection{Wind Turbine Dataset: Validation on a Complex Linear System}

To show that the core limitation is non-linearity rather than dimensional complexity, we use the IEA 15\, MW reference wind turbine dataset \citep{Pedersen2024}. Simulations yield high-dimensional responses for a system with hundreds of DOFs that behaves predominantly linearly in different operational conditions.

\begin{figure}
     \centering
     \begin{subfigure}[b]{0.3\textwidth}
         \centering
         \includegraphics[width=\textwidth]{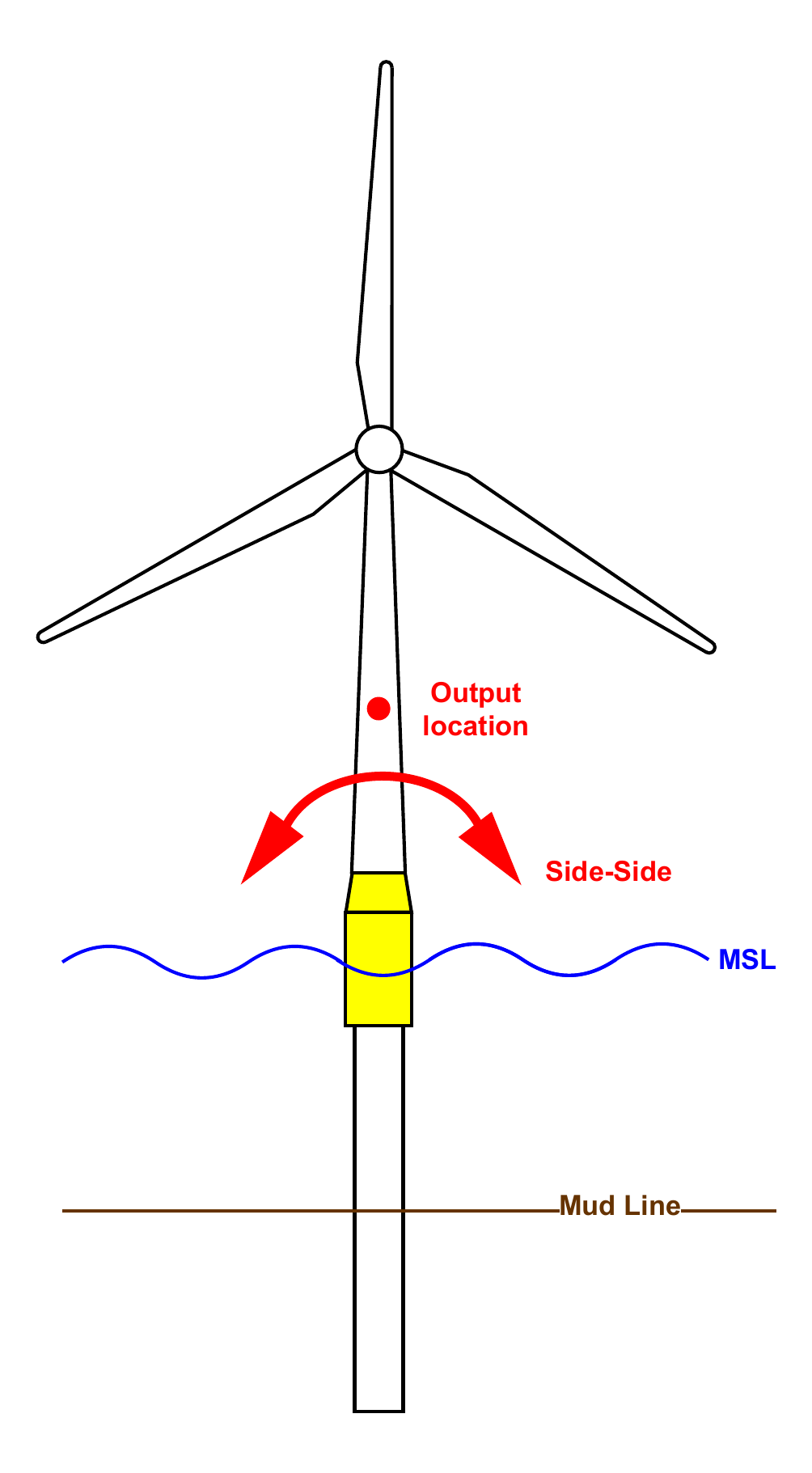}
         \caption{Side–side}
         \label{fig:WT_side-side}
     \end{subfigure}
     \begin{subfigure}[b]{0.29\textwidth}
         \centering
         \includegraphics[width=\textwidth]{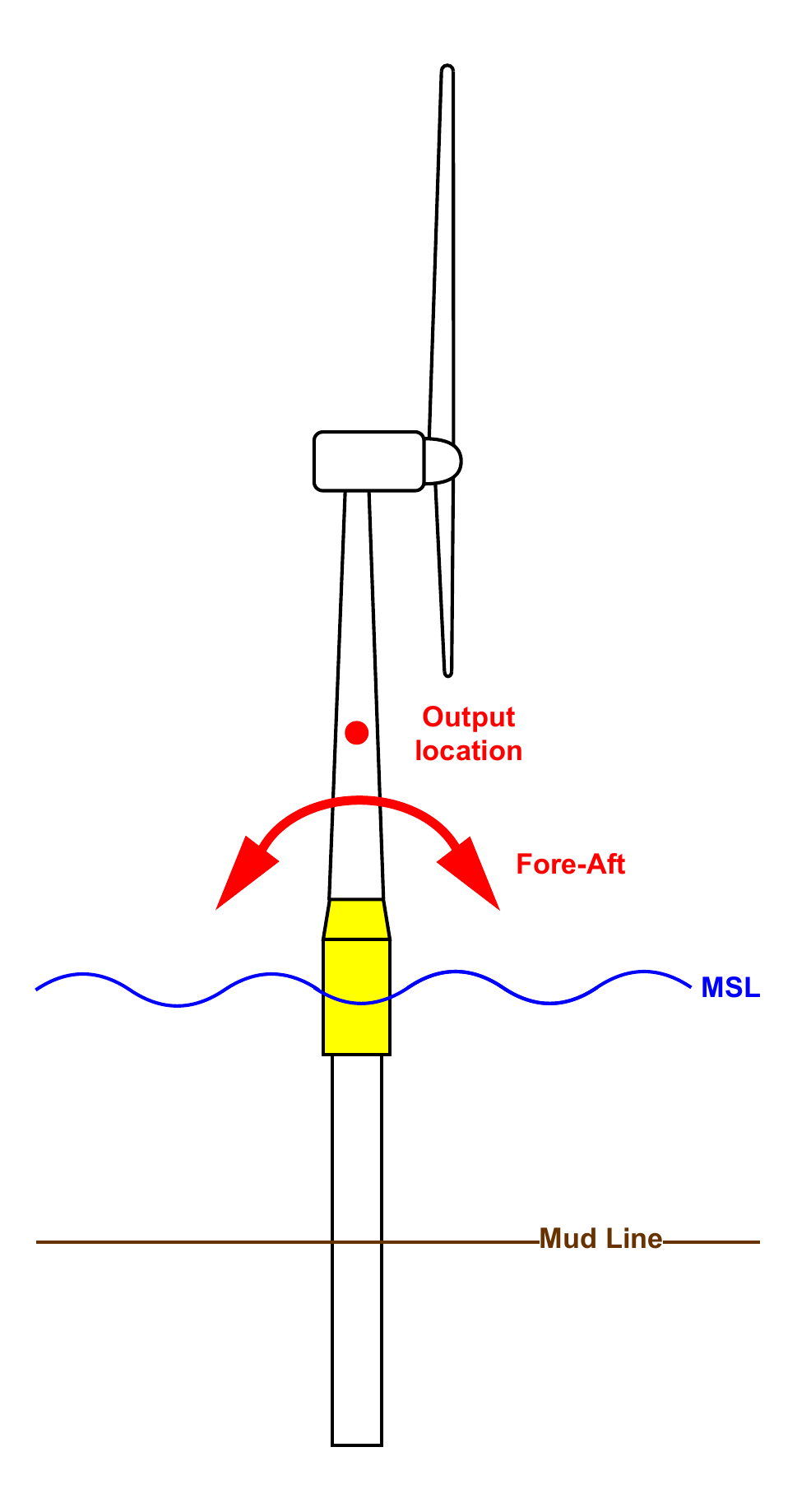}
         \caption{Fore–aft}
         \label{fig:WT_fore-aft}
     \end{subfigure}
     \caption{Offshore wind turbine schematic, moment directions, and output locations above the MSL.}
     \label{fig:OWT_FA_SS}
\end{figure}

We use two DLCs for training, and one for testing:

\begin{itemize}
\item \textbf{DLC~1.2:} Normal power production with active pitch and torque control.
\item \textbf{DLC~6.4:} Parked/idling with feathered blades.
\item \textbf{DLC~4.1:} Shut down situation, when the turbine used mechanical and/or aerodynamic brakes to stop electricity generation suddenly.
\end{itemize}
Each simulation provides 600\,s time series at 10\, Hz for multiple channels, including fore–aft and side–side bending moments at several tower heights. Figure \ref{fig:OWT_FA_SS} shows the schematic drawing of a fixed-bottom offshore wind turbine on a monopile foundation, and the bending moment directions. For the sake of space, in the results section we only present the results at 75m Mean Sea Level (MSL) location, which is indicated in the Figure \ref{fig:OWT_FA_SS}.

\subsection{Data Preprocessing Pipeline}

All signals are normalized prior to training. For the 2DOF data, forcing and displacements are standardized (mean/standard deviation) using training-set statistics and denormalized for metric computations that require physical units.

For the wind turbine, inputs comprise six SCADA channels:
\begin{itemize}
\item Wind speed [m/s] (environmental forcing)
\item Rotor speed [rad/s] (rotational dynamics)
\item Electrical power [W] (operational state)
\item Blade pitch angle [deg] (control response)
\item Tower-top acceleration $x$ [m/s$^2$] (side–side)
\item Tower-top acceleration $y$ [m/s$^2$] (fore–aft)
\end{itemize}
A seventh input encodes normalized height along the tower (0 at seabed to 174.6\,m at tower top), scaled to $[0,1]$. Outputs are 12 channels: fore–aft ($M_x$) and side–side ($M_y$) bending moments at six heights.

Because of disparate channel scales (e.g., watts vs. m/s), we apply min–max normalization to all inputs/outputs, computed globally across the training portions of DLC~1.2 and DLC~6.4, and reuse those parameters at inference. Training and testing include both DLCs to span the operational envelope.
 
\FloatBarrier

\section{Results and Discussion}
\label{sec:result}

In this section, we provide the results for the tests we ran on the trained FNO. We start with the 2DOF system and show the effect of the dataset size used to train the FNO on the test results. Based on the results of these tests, we choose the best-trained model for each of the 12 cases and examine their performance. We then study the effects of input-force changes, where we test the trained FNOs on forces that are different from those used during training. This section continues with the IEA 15MW turbine results, and ends with a discussion about the FNOs discretization error.

\subsection{Dataset size and FNO model Performance}
\label{sec:res_dataset_change}

\begin{figure}
    \centering
    \includegraphics[width=1\linewidth]{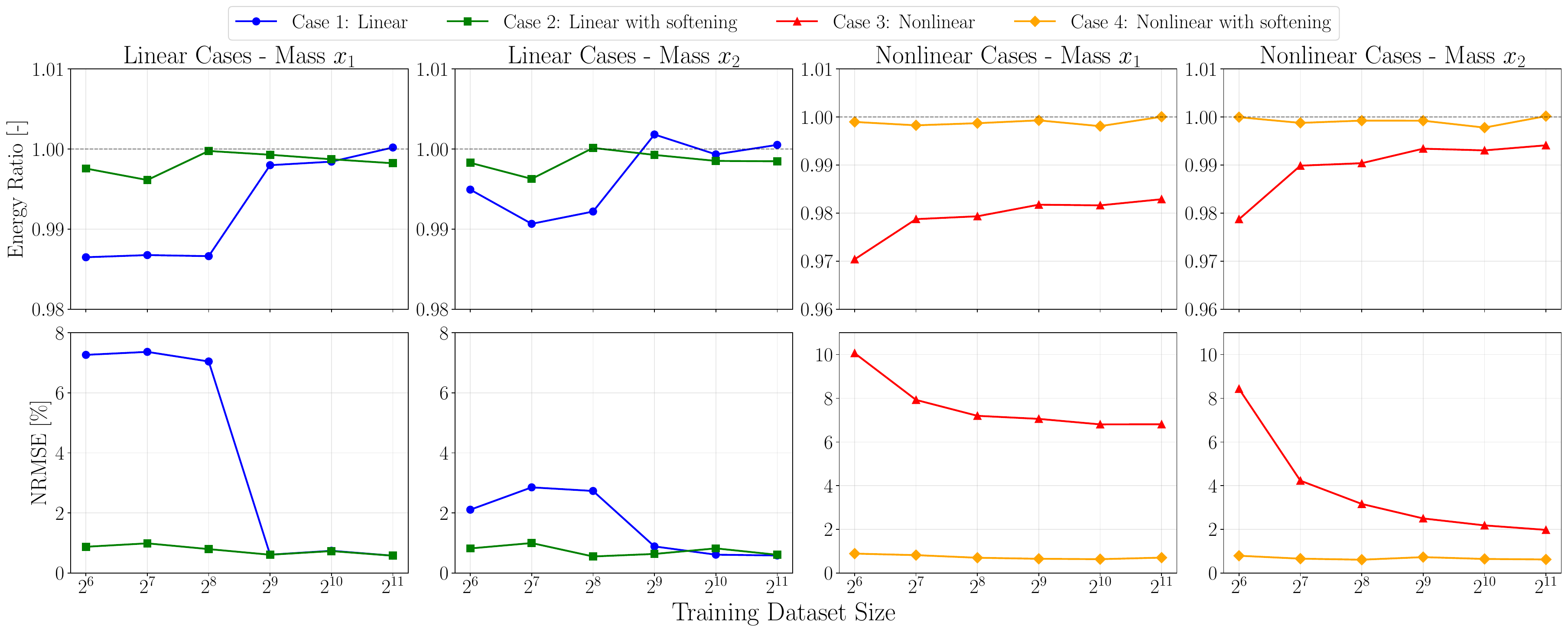}
    \caption{The effect of the dataset size on the low frequency configuration error. The energy ratio scale on the y-axis is between 0.96 and 1.01, while the NRMSE is below 10\%. By increasing the training dataset size, the linear cases show lower error, while the softening configuration in both linear and non-linear cases encloses smaller error due to FNO low-frequency bias.}
    \label{fig:db_size_lowfreq}
\end{figure}

\begin{figure}
    \centering
    \includegraphics[width=1\linewidth]{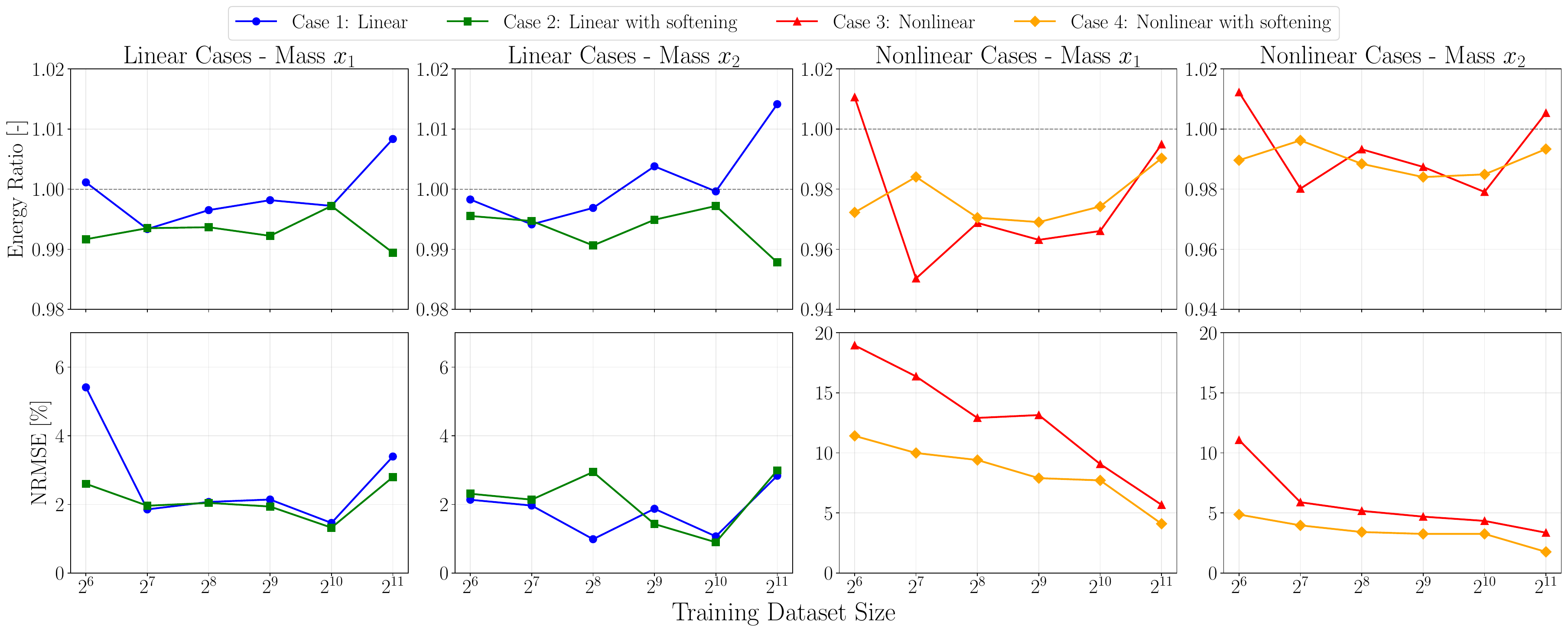}
    \caption{The effect of the dataset size on the high frequency configuration. The energy ratio scale on the y-axis is between 0.94 and 1.02, while the NRMSE is below 20\%. For the linear cases, for both error metrics, the error is low, while the increase in the training dataset size does not provide a meaningful pattern. For the non-linear cases, the increase in the training dataset size decreases the NRMSE, and the non-linear case with softening performs better due to the reduction in the system eigenfrequency.}
    \label{fig:db_size_highfreq}
\end{figure}

\begin{figure}
    \centering
    \includegraphics[width=1\linewidth]{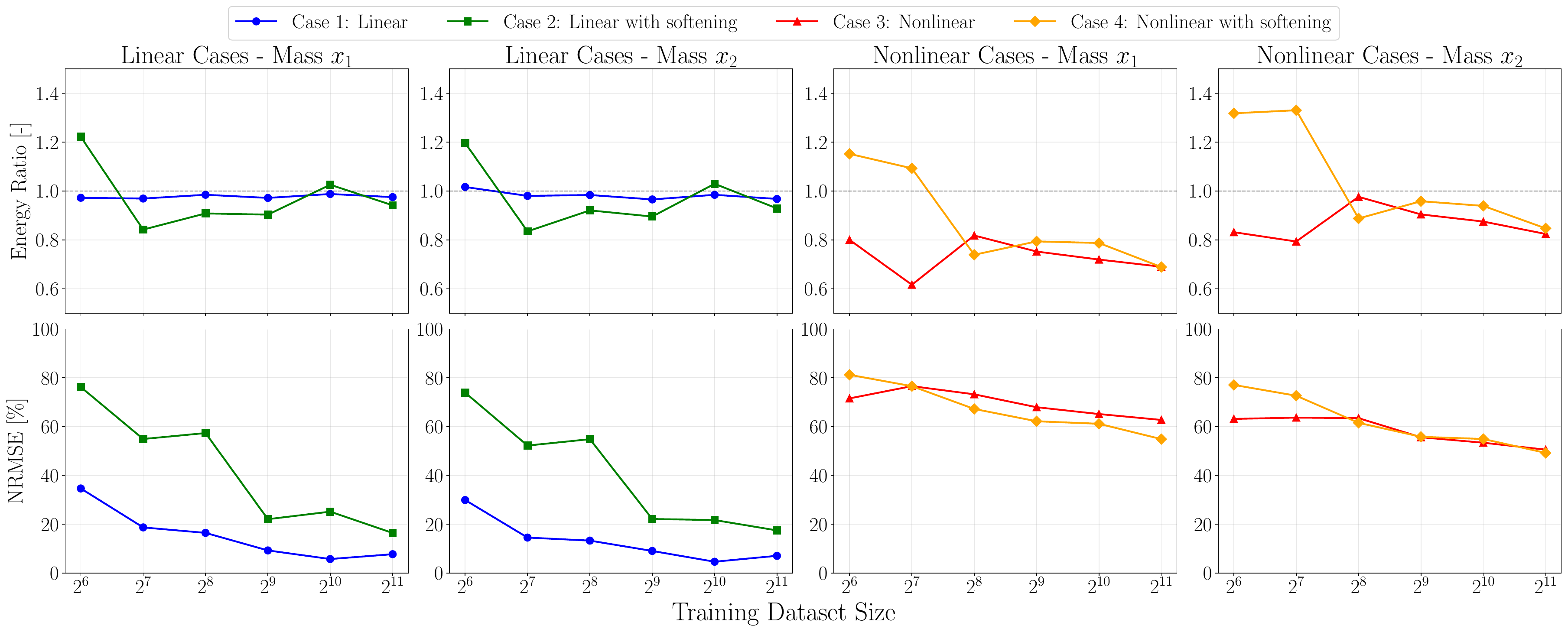}
    \caption{The effect of the dataset size on the broadband configuration. The energy ratio scale on the y-axis is between 0.5 and 1.5, while the NRMSE is between 0\% to 100\%. The linear cases show better results, with improvement in the error for the larger training dataset. The non-linear cases show marginal improvement in both error metrics.}
    \label{fig:db_size_broadband}
\end{figure}

We trained the FNOs for each case in Table \ref{tab:cases}, for the frequency configuration in Table \ref{tab:freq_conf}. For each of these cases, we trained the FNO on 6 dataset sizes mentioned in Section \ref{sec:data}. This means, in total, we trained 72 models for this study. The testing dataset, depending on the configuration, is created utilizing 2048 test samples. For each test, we loaded the trained model and then tested it on the 2048 samples, calculating the average error over all 2048 samples. The average error over 2048 samples is presented in Figures \ref{fig:db_size_lowfreq} to \ref{fig:db_size_broadband}.

When we train the FNO on a low-frequency dataset, Figure \ref{fig:db_size_lowfreq} shows interesting trends. For the linear cases, the FNO—even when trained on a small dataset—can predict the displacement for both masses with low error. However, as we move to the non-linear cases, the story changes. The FNO shows improved accuracy for Case 3 with larger datasets, though the improvement appears to flatten,  with increasing dataset size. This shows the FNO is hitting a bound, and the increase in dataset size does not improve performance. However, for the non-linear case with softening spring (Case 4)  the predictions are good at all the dataset sizes. This is due to the reduction in the frequency range of the system as the softening happens. This aligns with the literature on the FNO low-frequency bias \citep{oommenIntegratingNeuralOperators2025}. In all cases, the energy in the prediction signal is close to the actual signal. As the energy in the system comes from the input force—and for this case, the force input is at low frequency—the FNO prediction can capture the energy in the system response.

When we move to the system where the input force consists of high-frequency content (1\, Hz and 2\, Hz, Fig. \ref{fig:db_size_highfreq}), the FNO prediction on the testing dataset shows different behavior with respect to the dataset size. To start with, the error for all the cases is higher; still, the linear cases can be predicted with high accuracy. However, when we move to the non-linear cases, the FNO predictions have higher errors than the linear cases across the board. The error decreases with the increase in the dataset size used for training. Interestingly, the error for the non-linear case with softening is smaller, due to the reduction in the system eigenfrequency, which helps the FNO in prediction. From Figures \ref{fig:db_size_lowfreq} and \ref{fig:db_size_highfreq} one may conclude that FNO prediction is sensitive to the frequency content of the system excitation.

The frequency content of the forcing function fundamentally determines FNO learning dynamics through interaction with the spectral bias of the architecture. For low-frequency forcing, the FNO's inherent preference for low frequencies \citep{oommenIntegratingNeuralOperators2025} creates favorable learning conditions even with minimal training data. Linear systems maintain energy ratios $\approx 1.0$ across all dataset sizes because the operator mapping frequencies lie within the natural approximation space of FNOs. The sudden accuracy improvement for Case 1 at $n = 2^9$ samples corresponds to crossing the threshold where the number of training samples exceeds the effective dimensionality of the low-frequency subspace, enabling exact interpolation without aliasing errors \citep{kovachki2024data}.

For high-frequency forcing, the recovery pattern in the energy ratio is due to spectral truncation effects and spectral diversity in the training data. Initially, with $n < 2^8$ samples, the FNO learns an incomplete representation where high-frequency modes are aliased into lower frequencies, causing energy dissipation quantified by $E_{out}/E_{in} < 1$. The energy ratio reaches its minimum near $n = 2^7$ because the model has sufficient capacity to memorize training data but insufficient constraints to preserve energy conservation in the learned operator. Recovery occurs for $n > 2^9$ when the dataset provides enough spectral diversity to resolve the Nyquist criterion for the non-linear harmonics. 

In Figures \ref{fig:db_size_lowfreq} and \ref{fig:db_size_highfreq}, the contrast between frequency regimes reveals that FNO limitations are not universal but frequency dependent. Low-frequency problems benefit from the inductive bias of the architecture, where the truncated Fourier modes $k \leq k_{max}$ naturally capture the solution space. High-frequency problems suffer from the ``spectral bottleneck'' where critical information exists in modes $k > k_{max}$ that are systematically discarded \citep{li2021fourier}. The non-monotonic behavior observed in non-linear cases exhibits three distinct regimes: (i) with limited training data ($n < 2^7$), the model undergoes mode collapse, learning only dominant frequencies while failing to capture the complete spectral content; (ii) at intermediate dataset sizes ($2^7 \leq n \leq 2^9$), partial learning occurs with systematic energy dissipation due to insufficient constraints for proper operator approximation; (iii) with sufficient data ($n > 2^9$), the model achieves accurate spectral representation, as gradient descent naturally converges to solutions with good generalization properties.

In Figure \ref{fig:db_size_broadband}, the systematic performance degradation from linear (Cases 1–2) to non-linear systems (Cases 3–4) indicates fundamental limitations in how FNOs handle non-linear dynamics under stochastic forcing. While linear systems achieve near-unity energy preservation (RMS ratio $\approx 1.0$) and PSD shape errors below 20\% with as few as $2^8$ training samples, non-linear systems exhibit persistent energy dissipation ($\approx 30\%$ loss) and PSD errors exceeding 60\% regardless of training dataset size. This disparity stems from spectral truncation of FNOs, which discards high-frequency information precisely where non-linear systems cascade energy through mode coupling. The broad range of frequencies of the forcing function intensifies this issue in non-linear cases. The cubic non-linearity $(k_3(x_1-x_2)^3)$ generates harmonics beyond the FNO's retained modes, creating an artificial energy sink that additional training data cannot overcome. The time-varying stiffness (softening) introduces additional complexity that slightly degrades performance in both linear and non-linear cases. However, its impact is less dominant than the effect of non-linearity. Notably, the data requirements for modeling differs dramatically. The linear systems plateau at acceptable performance with $2^8$ samples, while non-linear systems show minimal improvement even at $2^{11}$ samples, suggesting the error is likley due to the FNO architecture rather than the statistics of the training data.

The contrasting performance between stochastic forcing (broadband frequencies in Figure \ref{fig:db_size_broadband}) and fixed-frequency forcing (Figures \ref{fig:db_size_lowfreq} and \ref{fig:db_size_highfreq}) demonstrates that FNOs struggle with the response \textit{complexity} of non-linear systems under random excitation, rather than harmonic excitations alone. When forced at fixed frequencies, all cases, including non-linear systems, achieve near-unity energy preservation and PSD errors below 10\%, showing that FNOs can accurately capture deterministic non-linear dynamics for deterministic excitation. However, under stochastic forcing with frequencies randomly sampled from distributions, non-linear cases exhibit persistent 30–50\% energy loss and PSD errors remaining above 60\% even with $2^{11}$ training samples. Simultaneously, linear cases maintain acceptable performance and improve with larger training datasets.

This performance difference suggests the challenge lies in learning the response of the non-linear system across the forcing-frequency distribution space. For linear systems, the frequency response function remains fixed regardless of input amplitude, allowing FNOs to efficiently learn the mapping through superposition, provided there are large datasets for training. Non-linear systems, however, exhibit amplitude-dependent frequency responses where the cubic terms $k_3(x_1-x_2)^3$ create different spectral patterns for each realization of the random forcing. The FNO must therefore learn not just a simple mapping to one frequency but rather an ensemble of mappings conditioned on the input spectrum—a significantly more complex learning task that appears to exceed the capacity of standard FNO architectures with fixed mode truncation. This explains why increasing training data provides minimal improvement: the issue is not inadequate data but rather the architectural inability to represent the full spectral content required for non-linear systems under broadband stochastic excitation.

\subsection{Reference case LSTM}

\begin{figure}
    \centering
    \includegraphics[width=1\linewidth]{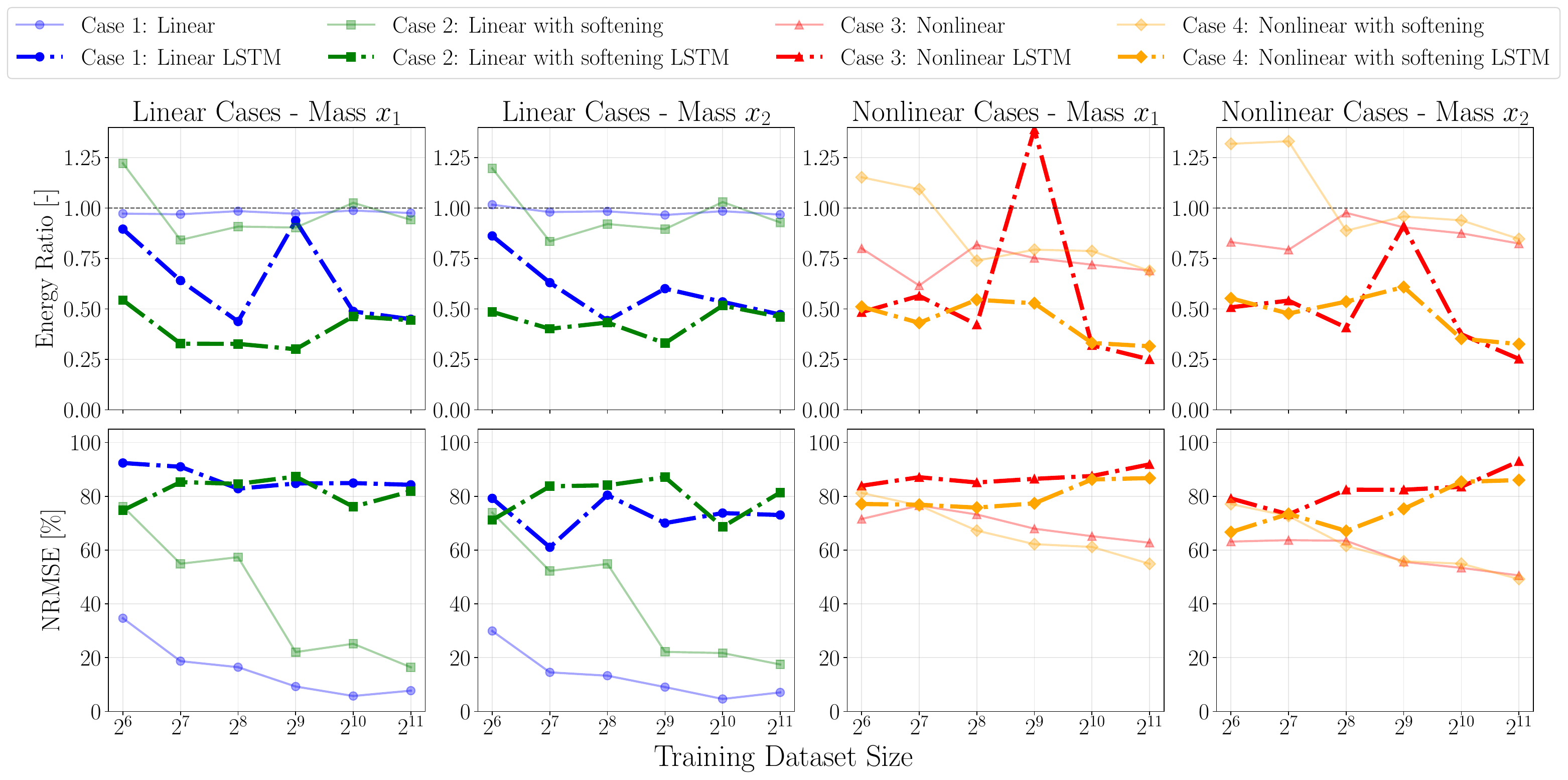}
    \caption{The effect of the dataset size on the broadband configuration for the LSTM model. The broadband configuration results (Figure \ref{fig:db_size_broadband}) are kept in the figure, but faded for reference. The LSTM does not show any meaningful improvement in the results when the training dataset size increases. Across all the cases and for both error metrics, LSTM fails to predict the output.}
    \label{fig:db_size_broadband_lstm}
\end{figure}

In this study, to have a comparison base for the FNOs, we trained a model consisting of a two-layer LSTM combined with fully connected layers with 128 and 64 hidden units, on the broadband frequency excitation for all the cases. This provided a base to understand and study how FNO performs against traditional sequential data or time-series models.  The LSTM here is tested on the 2048 samples that were not used in the training of any of the models. Figure \ref{fig:db_size_broadband_lstm} shows the evaluation of the trained LSTM model on the test results when it is trained on different dataset sizes.

The comparative analysis between LSTM and FNO architectures demonstrates fundamental differences in their ability to learn and predict the dynamics of the 2DOF system across varying complexity levels. The most notable observation is the systematic energy dissipation exhibited by the LSTM models, with RMS energy ratios consistently hovering around 0.5 across all cases and dataset sizes, indicating that the predicted responses contain only half the energy of the true system dynamics. This energy loss remains constant regardless of training dataset size, suggesting an architectural limitation rather than a data scarcity issue. In contrast, the FNO demonstrates substantially superior energy preservation, particularly for the linear system (Case 1), where the energy ratio approaches one with increasing dataset size. Even for the more challenging non-linear and time-varying stiffness cases, FNO maintains energy ratios between 0.7 and 1.0, expressing a significant improvement over LSTM performance. This difference underlines the FNO's advantage in preserving global conservation properties through its spectral representation, whereas LSTM's sequential processing appears to introduce artificial numerical dissipation that cannot be overcome through increased training data alone.

In Figure \ref{fig:db_size_broadband_lstm}, the frequency-domain analysis further highlights the architectural advantages of the FNO for dynamical system modeling. The PSD shape error metrics show that FNO achieves significant spectral accuracy for the linear system, with errors dropping below 10\% when trained on 2048 samples or more, while LSTM maintains errors between 60–90\% regardless of dataset size. This performance gap can be attributed to the FNO's explicit operation in the frequency domain through the FFT, allowing it to directly learn spectral relationships and mode interactions. The data scaling behavior presents another crucial distinction.  FNO exhibits clear performance improvements with increasing dataset sizes across all metrics, following expected deep learning scaling laws, whereas LSTM shows negligible improvement beyond 64 training samples. This plateau in LSTM performance suggests that the architecture quickly reaches its limits for this problem class. Additionally, it demonstrates that LSTM is unable to leverage additional data to capture the complex temporal dependencies over the entire signal length.

The system complexity gradient from linear to non-linear with time-varying stiffness reveals interesting patterns in both FNO and LSTM capabilities (Figure \ref{fig:db_size_broadband_lstm}). While both models struggle with the non-linear cases (Cases 3 and 4), the energy degradation patterns differ significantly. FNO maintains reasonable performance even for the most complex case (non-linear with softening), with PSD errors around 50–60\% and energy ratios above 0.7, suggesting that it captures the essential dynamics despite imperfect accuracy. However, LSTM shows uniformly poor performance across all complexity levels, with only marginal differences between linear and non-linear cases. This indicates a fundamental inability to distinguish between different dynamical regimes. These results strongly support the use of FNO architectures for the long-term integration of dynamical systems, especially when it is crucial to maintain accuracy and conserve energy. In contrast, LSTM architectures may require significant modifications, such as utilizing physics-informed loss functions, to achieve a comparable level of performance on these problems. We conducted similar tests on the low- and high-frequency datasets using LSTM, with no meaningful changes in the results. We need to keep in mind that even for the low- and high-frequency datasets, the phase and amplitude of the excitation come from a uniform distribution, which pushes LSTM to its limits.

\subsection{Effect of Excitation Force Choice in FNO}

Based on the results in Section \ref{sec:res_dataset_change}, we choose the best model from the 72 cases. Then, these FNOs were tested against three excitation forces. We ran the tests on all the configurations in Table \ref{tab:freq_conf} and cases in Table \ref{tab:cases}. First, we tested the trained FNOs against a chirp force signal. This signal has a changing frequency content from 0.01\, Hz to 2\, Hz over 200\,s of the signal. This means the 2DOF gets a wide range of frequency excitation. Then, we tested the trained models against an impulse signal, as the frequency content of an impulse signal excitation spans a full frequency range. Finally, we tested the trained FNOs against sudden changes in the force using a step signal.

In the case of the wind turbine, the model was trained on the operational and idling DLCs. Then we tested the trained FNO on the wind turbine simulation output for a shutdown case. The shutdown case is when the turbine suddenly stops generating electricity by pitching out the blades (aerodynamic brake) and/or using a mechanical brake, depending on the wind turbine configuration.

As we are comparing response to single inputs and not an ensemble,  we use different error metrics than in Section \ref{sec:res_dataset_change}. We choose instead Energy Ratio and Spectral Coherence as the error metrics. Both of these error metrics are explained in Section \ref{sec:meth_error_metric}. 

\subsubsection{2DOF System}

\begin{figure}
    \centering
    \includegraphics[width=1\linewidth]{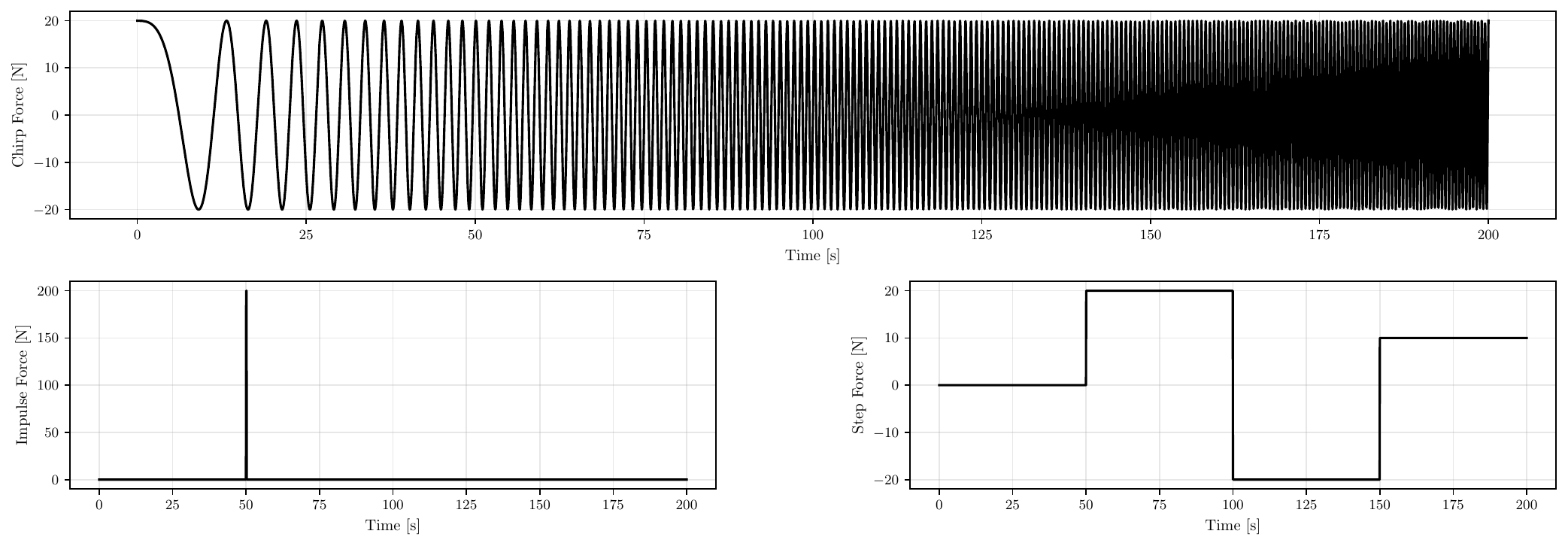}
    \caption{The input forces used for tests. From top to bottom and  left to right, the chirp force input, when the frequency of excitation changes with time, the impulse force when the system is excited with a sudden force at 50s, and the step force input ,where there are sudden changes in the force.  }
    \label{fig:input_force}
\end{figure}

In the following, we present the test results on the trained FNOs for the 2DOF system. Figure \ref{fig:input_force} shows these test input forces as explained before.

\begin{landscape} 
\begin{table*}
\centering
\caption{FNO Performance Metrics Across Different Frequency Domains}
\label{tab:fno_metrics}
\setlength{\tabcolsep}{2.5pt} 
\renewcommand{\arraystretch}{1.08}
\begin{adjustbox}{width=\linewidth}
\begin{tabular}{@{}ll*{24}{c}@{}}
\toprule
\multirow{3}{*}{\textbf{Case}} & \multirow{3}{*}{\textbf{Var}} &
\multicolumn{6}{c}{\textbf{Low Frequency}} &
\multicolumn{6}{c}{\textbf{High Frequency}} &
\multicolumn{6}{c}{\textbf{Broadband}} &
\multicolumn{6}{c}{\textbf{Broadband w/ Spectrogram Loss}} \\
\cmidrule(lr){3-8}\cmidrule(lr){9-14}\cmidrule(lr){15-20}\cmidrule(lr){21-26}
& & \multicolumn{2}{c}{\textbf{Chirp}} & \multicolumn{2}{c}{\textbf{Impulse}} & \multicolumn{2}{c}{\textbf{Sudden}} &
      \multicolumn{2}{c}{\textbf{Chirp}} & \multicolumn{2}{c}{\textbf{Impulse}} & \multicolumn{2}{c}{\textbf{Sudden}} &
      \multicolumn{2}{c}{\textbf{Chirp}} & \multicolumn{2}{c}{\textbf{Impulse}} & \multicolumn{2}{c}{\textbf{Sudden}} &
      \multicolumn{2}{c}{\textbf{Chirp}} & \multicolumn{2}{c}{\textbf{Impulse}} & \multicolumn{2}{c}{\textbf{Sudden}} \\
\cmidrule(lr){3-4}\cmidrule(lr){5-6}\cmidrule(lr){7-8}
\cmidrule(lr){9-10}\cmidrule(lr){11-12}\cmidrule(lr){13-14}
\cmidrule(lr){15-16}\cmidrule(lr){17-18}\cmidrule(lr){19-20}
\cmidrule(lr){21-22}\cmidrule(lr){23-24}\cmidrule(lr){25-26}
& & \textbf{ER} & \textbf{SC(\%)} & \textbf{ER} & \textbf{SC(\%)} & \textbf{ER} & \textbf{SC(\%)} &
      \textbf{ER} & \textbf{SC(\%)} & \textbf{ER} & \textbf{SC(\%)} & \textbf{ER} & \textbf{SC(\%)} &
      \textbf{ER} & \textbf{SC(\%)} & \textbf{ER} & \textbf{SC(\%)} & \textbf{ER} & \textbf{SC(\%)} &
      \textbf{ER} & \textbf{SC(\%)} & \textbf{ER} & \textbf{SC(\%)} & \textbf{ER} & \textbf{SC(\%)} \\
\midrule
\multirow{2}{*}{\textbf{Linear}} 
& $x_1$ & 0.172 & 44.2 & 0.130 & 98.7 & 0.340 & 92.9 & 0.041 & 44.8 & 0.063 & 94.1 & 0.206 & 85.2 & 0.921 & 49.7 & 0.783 & 97.3 & 1.086 & 87.5 & 0.893 & 44.8 & 0.644 & 91.6 & 0.632 & 91.8 \\
& $x_2$ & 0.257 & 41.9 & 0.186 & 97.0 & 0.340 & 83.8 & 0.046 & 36.9 & 0.093 & 92.6 & 0.269 & 65.2 & 0.915 & 42.0 & 0.708 & 91.6 & 1.050 & 48.7 & 0.923 & 44.4 & 0.597 & 93.6 & 0.495 & 70.7 \\
\midrule
\multirow{2}{*}{\textbf{Linear Softening}} 
& $x_1$ & 0.569 & 48.3 & 0.422 & 99.0 & 0.668 & 96.6 & 0.041 & 46.5 & 0.161 & 97.5 & 0.038 & 96.5 & 0.998 & 49.3 & 0.756 & 93.7 & 0.838 & 90.3 & 0.991 & 42.8 & 0.866 & 92.5 & 0.938 & 38.5 \\
& $x_2$ & 0.606 & 49.6 & 0.450 & 94.9 & 0.632 & 66.3 & 0.043 & 45.3 & 0.131 & 92.6 & 0.035 & 50.6 & 0.997 & 48.7 & 0.738 & 87.7 & 0.855 & 55.2 & 1.029 & 43.7 & 0.892 & 87.5 & 0.978 & 27.4 \\
\midrule
\multirow{2}{*}{\textbf{non-linear}} 
& $x_1$ & 0.175 & 67.6 & 0.242 & 94.6 & 0.364 & 96.8 & 0.049 & 43.9 & 0.074 & 95.4 & 0.146 & 56.9 & 0.335 & 15.9 & 0.248 & 67.6 & 0.558 & 16.5 & 0.300 & 47.7 & 0.202 & 87.9 & 0.386 & 94.1 \\
& $x_2$ & 0.248 & 63.9 & 0.384 & 87.2 & 0.296 & 77.4 & 0.060 & 54.2 & 0.108 & 87.5 & 0.189 & 66.2 & 0.446 & 16.4 & 0.385 & 85.8 & 0.704 & 56.7 & 0.403 & 38.8 & 0.245 & 81.7 & 0.356 & 73.2 \\
\midrule
\multirow{2}{*}{\textbf{non-linear Softening}} 
& $x_1$ & 0.406 & 68.7 & 0.386 & 96.6 & 0.670 & 95.0 & 0.062 & 55.8 & 0.078 & 93.0 & 0.049 & 82.6 & 0.908 & 64.0 & 0.670 & 94.7 & 1.007 & 89.3 & 0.732 & 44.7 & 0.553 & 92.7 & 0.527 & 67.7 \\
& $x_2$ & 0.420 & 70.7 & 0.420 & 91.0 & 0.643 & 63.6 & 0.071 & 59.4 & 0.090 & 87.1 & 0.049 & 61.3 & 1.031 & 62.7 & 0.755 & 85.4 & 1.013 & 51.1 & 0.846 & 48.2 & 0.623 & 83.5 & 0.531 & 23.4 \\
\bottomrule
\end{tabular}
\end{adjustbox}
\vspace{2mm}
\footnotesize
\textbf{Note:} ER = Energy Ratio, SC = Spectral Coherence. The table presents FNO performance metrics across four frequency categories: Low Frequency, High Frequency, Broadband, and Broadband with Spectrogram Loss. Each domain is tested with three input signals (Chirp, Impulse, Sudden Changes) and evaluated on four system cases (Linear, Linear Softening, Non-linear, Non-linear Softening) for two state variables ($x_1$ and $x_2$).
\end{table*}
\end{landscape} 

We tested the trained FNOs on the excitation force signal in Figure \ref{fig:input_force}, and calculated the Energy Ratio and Spectral Coherence as the error metrics for each test. Table \ref{tab:fno_metrics} presents the results of these tests. Each row in Table \ref{tab:fno_metrics} presents the test results for each system case; the columns present the frequency configuration that each system is trained on.

Our results in Table \ref{tab:fno_metrics} demonstrate a fundamental limitation of FNOs when applied to non-linear dynamical systems: significant energy dissipation when the test signal frequency content differs from the training signal. This phenomenon is most dramatic in the high-frequency–trained models, which exhibit significant energy loss with ratios dropping to 0.041–0.206 for mass 1 displacement $x_1$ across all test cases  (Table \ref{tab:fno_metrics}, column 2). The energy dissipation is particularly severe for non-linear systems (Cases 3–4), for both the low-frequency– and high-frequency–trained models (Table \ref{tab:fno_metrics} columns 1 and 2). 
Interestingly, although poorly overall, both models show less error moving from Case 3 to Case 4 as softening happens in Case 4. This again shows how the low-frequency bias in FNO can affect the results. This is more visible in the chirp test and sudden force change, where the energy ratio is more than doubled. Regardless of the training configuration and the case, the chirp test results have medium to low spectral coherence. This can be due to the sweeping frequency that the chirp case imposes on the system, which is a non-stationary signal. When spectral coherence is computed using the Welch method, this sweeping frequency affects the output.

The pattern is clear: FNOs trained on a limited frequency band (Table \ref{tab:fno_metrics}, columns 1 and 2) act as unintended frequency-selective filters. When encountering out-of-band signals, they artificially dampen the response rather than learning the true system dynamics. This is evident in the chirp tests, which sweep from 0.01–2\, Hz and thus probe the full frequency range. The high-frequency–trained model, having never seen low-frequency dynamics during training, essentially suppresses these components, leading to the observed energy loss.

An intriguing finding is the decoupling between spectral coherence and energy preservation. Even when energy ratios drop below 0.1, spectral coherence often remains high for linear and non-linear systems. This indicates that FNOs preserve the frequency-content distribution while failing to capture amplitude dynamics correctly. The model learns the shape of the frequency response but not its magnitude.

The non-linear cases (3–4) show dramatically worse energy preservation than the linear cases, with the effect most pronounced for broadband training (Table \ref{tab:fno_metrics}, column 3). This likely roots in the frequency coupling. The frequency coupling inherent in non-linear systems (the cubic stiffness term generates harmonics and intermodulation products) extends beyond the training frequency range. Representing this non-linear energy transfer between frequency bands can not be done with an FNO trained on specific frequencies.


The broadband-trained models show intermediate performance, avoiding total failure but achieving lower peak accuracy (Table \ref{tab:fno_metrics}, column 3). The combination of energy ratios and spectral coherence suggests the model learns a representation that partially captures dynamics across the frequency spectrum, but it is not good at any one. This shows a necessary trade-off between robustness and accuracy. 

The energy dissipation mechanism can be understood through the FNO architecture. The global Fourier transform in each layer truncates high-frequency modes (we retain only 1024 modes), and this truncation accumulates through the network depth. When the model faces frequencies outside its training distribution, the learned weights effectively act as a dissipative filter.
This is fundamentally different from traditional numerical methods, which may have truncation errors but generally preserve energy structure, especially for conservative integrators.

\subsubsection{IEA 15MW turbine stop}

We initially trained the FNO on the operational (DLC12) and idling (DLC64) conditions combined. This gave the FNO a wide spectrum of inputs. The normal operation and idling simulation results were divided into training and test cases. We trained the FNO on 474 normal operational simulation data and 172 idling simulation data. We separated 120 normal operation simulation data and 20 idling simulation data for testing to ensure there is no data leakage. The test results for DLC12 and DLC64 are presented in Table \ref{tab:wt_test}. For the sake of space, we only present one output at 75\,m MSL as shown in Figure \ref{fig:OWT_FA_SS}.

\begin{table}
\centering
\caption{FNO Error Metrics - Location 75m MSL}
\begin{tabular}{l cc cc cc}
\hline
 & \multicolumn{2}{c}{DLC12} & \multicolumn{2}{c}{DLC64} & \multicolumn{2}{c}{DLC41} \\
 \cmidrule(lr){2-3} \cmidrule(lr){4-5} \cmidrule(lr){6-7}
Component & ER & SC (\%) & ER & SC (\%) & ER & SC (\%) \\
\hline
M\textsubscript{FA} & 1.005 & 58.6 & 1.001 & 65.8 & 1.017 & 63.2 \\
M\textsubscript{SS} & 0.991 & 67.3 & 0.996 & 53.9 & 1.340 & 68.4 \\
\hline
\end{tabular}
\label{tab:wt_test}
\end{table}

The FNO can predict the moment time series in both FA and SS directions with high accuracy for both cases. Interestingly, for both cases, the FNO has a high energy ratio but medium spectral coherence. An OWT has mainly two sources of excitation: turbulent (unsteady) wind and waves. The turbulent wind frequency spectrum is wide and excites the structure similarly to the broadband frequency configuration \citep{mannWindFieldSimulation1998a}. In contrast, the wave frequency spectrum is concentrated at one main frequency \citep{hasselmannMeasurementsWindWaveGrowth}, which is similar to the 2DOF configuration of one dominant frequency. When the turbine is operational, the wind loads transferred via rotor thrust and torque are the dominant contributors to the FA and SS time series on the wind turbine tower and foundation. However, when the turbine is idling, the wave load is the dominant contributor to the FA and SS moments time series. These two phenomena are observable in the results presented in Table \ref{tab:wt_test}, and affect the FNO prediction. The FNO performs better for the idling cases, as the structure is excited with one low frequency, rather than the operational cases, where the structure is excited with a range of frequencies. One interesting observation here is the spectral coherence values. Although the energy ratio is high, the spectral coherence is not as high. This is due to the sensitivity of spectral coherence to even one or two time steps of phase difference \citep{priestley1981spectral}. 

To this point, we have tested this model trained on the operational and idling conditions on the same dataset. Another operational condition that a turbine would go through is shutdown (DLC41). There are different strategies for a wind turbine to suddenly stop when it is in operation. One of these methods is aerodynamic braking, which is sudden pitching out of the blades to stop them from capturing wind. This means the aerodynamic load on the structure suddenly stops. This is very similar to the test case when the 2DOF system faces an impulse or sudden load change. The last column in Table \ref{tab:wt_test} presents the results from the shutdown case test on the FNO trained on idling and operational load cases.

The performance of the FNO models on shutdown scenarios reveals fundamental insights into the transferability of learned dynamics across different operating conditions. The models, trained exclusively on operational and idling conditions, demonstrate an explicit ability to predict structural loads during shutdown events, which is a regime characterized by fundamentally different aerodynamic and control phenomena.

The energy ratio values, consistently approximating unity across multiple test cases, suggest that the FNO architecture successfully captures the intrinsic structural response characteristics of the wind turbine system. The slight variations observed in energy ratios, ranging from 1.017 to 1.340, reflect the model's capacity to approximate the total energy content of the response while occasionally exhibiting overshooting characteristics, notably apparent in the side–side moment predictions, where ratios exceed unity.

The spectral coherence values, ranging from approximately 63\% to 68\%, provide deeper insight into the nature of the FNOs predictions. The coherence values in this range show that while amplitude correlation remains significant, phase relationships indicate moderate decorrelation. The observed spectral coherence values suggest phase variations of approximately $\pm$30–45 degrees. These values align with the complexities of predicting transient aerodynamic phenomena during rapid deceleration.

From a practical engineering perspective, these results have significant implications. The FNOs' ability to predict load magnitudes accurately without exposure to shutdown conditions during training suggests that fundamental structural response patterns were learned by the FNOs. This finding supports the claim that FNOs learn underlying physical principles rather than simply mimicking condition-specific responses, provided the model is complex and linear. 

These findings suggest that the FNO framework successfully extracts transferable features from the training data that generalize to previously unseen operational conditions. The shutdown regime, characterized by rapid pitch actuation, aerodynamic instabilities during rotor deceleration, and transient structural responses, represents a significant distance from quasi-steady conditions present in operational and idling states. The fact that the FNO maintains energy-ratio accuracy near unity while achieving moderate spectral coherence demonstrates that it learns the fundamental system dynamics.

One may conclude that incorporating physics-informed constraints could potentially improve phase accuracy while maintaining the current amplitude prediction capability. The repeated shutdown cases indicate that the phase errors are likely systematic rather than random. This means they could be addressed through post-processing or changes to the FNO design that account for time delays in the interaction between aerodynamics and structure during transient events.

\subsection{Adapting Loss function with Spectrogram Loss}

\begin{figure}
    \centering
    \includegraphics[width=1\linewidth]{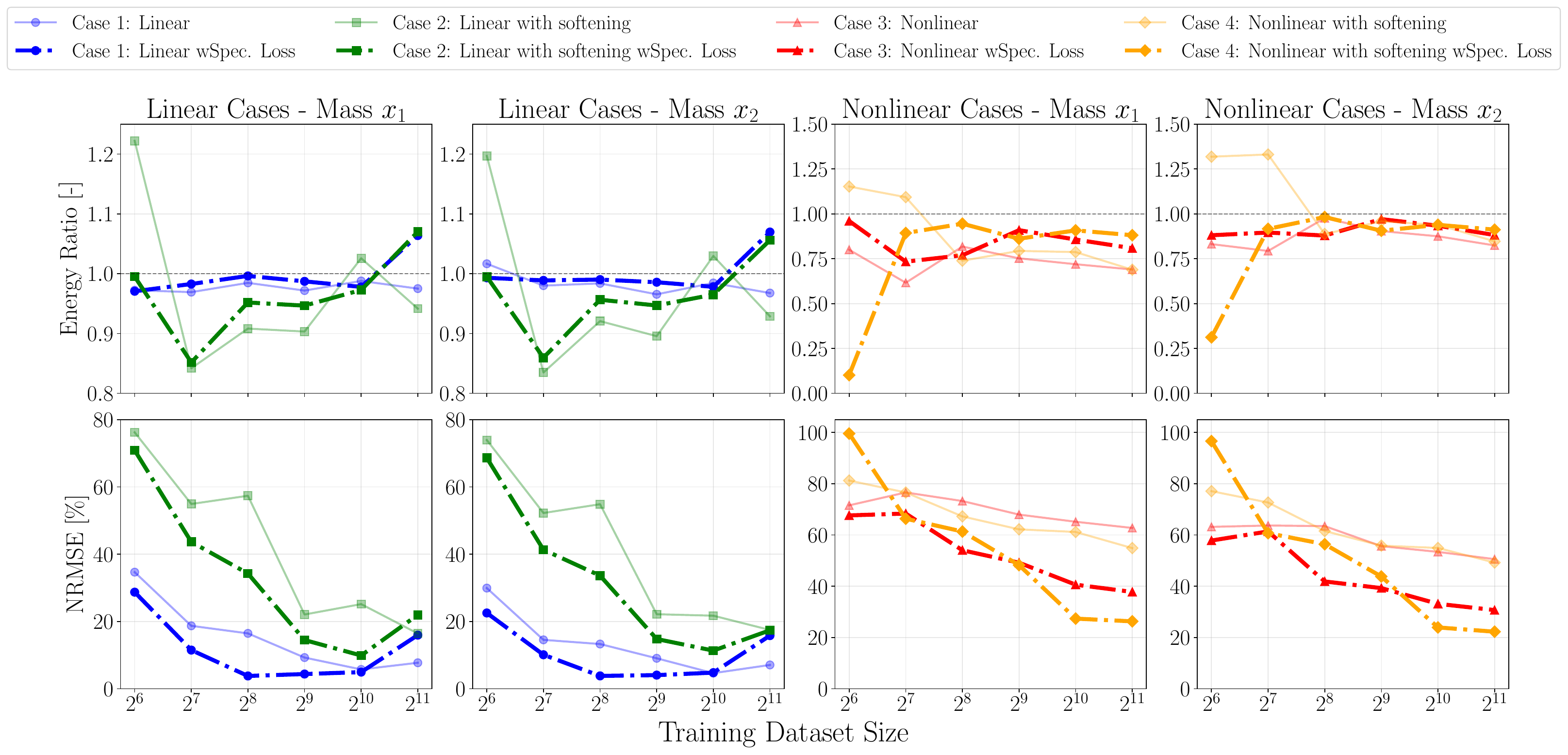}
    \caption{The broadband configuration with spectrogram loss showing the effect of dataset size on model performance. The broadband configuration results (Figure \ref{fig:db_size_broadband}) are kept in the figure, but faded for reference. Adding the Spectrogram Loss shows improvement in both error metrics, by decreasing the NRMSE and improving the energy ratio. The non-linear cases show a more significant improvement, both in the magnitude of the metrics and in the trajectory.}
    \label{fig:db_broadband_wSpecLoss}
\end{figure}

Throughout our tests, one of the main limitations of FNOs has been their bias towards lower frequencies. To address this fundamental challenge, we incorporated spectrogram loss into the training objective, enabling the model to maintain fidelity across the full frequency spectrum (Section \ref{sec:spectroloss}). Figure \ref{fig:db_broadband_wSpecLoss} shows the improvements achieved by FNO models trained with spectrogram loss compared to standard MSE-only training across different dataset sizes.

For the conservative mechanical systems, similar to what we studied here, which lack physical damping, the total energy in the time series should remain constant (Cases 1 and 3) or vary only with the time-dependent stiffness decay (Cases 2 and 4). We showed that the standard FNO training with MSE alone produces models that exhibit artificial numerical dissipation due to mode truncation (Figures \ref{fig:db_size_lowfreq}, \ref{fig:db_size_highfreq}, and \ref{fig:db_size_broadband}). In contrast, spectrogram-augmented training reaches energy ratios consistently near 1.0 across all cases for dataset sizes between $2^7$ and $2^9$ (Figure \ref{fig:db_broadband_wSpecLoss}). The addition of the spectrogram loss practically eliminates this non-physical energy loss in the data-efficient regime.

In Figure \ref{fig:db_broadband_wSpecLoss}, the frequency-domain metrics show the mechanism underlying these improvements. Without spectrogram loss, the standard FNO exhibits PSD shape errors averaging 30\% for the linear case, 55\% for linear softening, and 65–80\% for non-linear cases. Including spectrogram loss reduces these errors substantially across the training data range, by achieving minima below 10\% for the linear case at $2^9$ samples, approximately 20\% for linear softening, and 30–40\% for non-linear cases at larger dataset sizes. The improvement in spectral accuracy is consistent with the observed energy preservation. This increase in spectral accuracy, which means a more accurate frequency-domain representation, translates into conservation principles via Parseval's theorem \citep{oppenheim2010}.  

The spectrogram loss addresses two fundamental challenges simultaneously, particularly for non-linear and time-varying systems.

\begin{itemize}
    \item First, it captures time-varying spectral content arising from system non-linearities and time-dependent parameters. In non-linear systems (Cases 3 and 4), the cubic stiffness term generates higher harmonics through frequency mixing, while the exponentially decaying stiffness in Cases 2 and 4 causes systematic frequency modulation. The time–frequency localization of the STFT enables the loss function to track these evolving spectral features, penalizing models that fail to reproduce harmonic generation or frequency shifts. This capability is particularly visible in Case 2, where PSD errors drop from 70\% to below 20\% as the spectrogram loss captures the continuous frequency decrease accompanying stiffness degradation.
    \item Second, the spectrogram loss directly counteracts the well-documented spectral bias in neural networks and FNOs, which persists in FNOs despite their frequency-domain operations \citep{rahaman2019spectral,oommenIntegratingNeuralOperators2025}.Without spectral regularization, FNOs trained with MSE tend to bias towar smooth, low-frequency modes that minimize pointwise errors while filtering out high-frequency oscillatory components, which are essential for long-term accuracy. The spectrogram loss forces the neural network to maintain accuracy across all available frequencies by {\it redistributing the weights} to minimize errors across the spectrum.
\end{itemize}


The weighted loss function (Equation \eqref{eq:total_loss}) creates an optimization framework that follows principles established in physics-informed neural networks \citep{raissi2019physics}, where complementary constraints enforce different physical properties. The 80–20 weighting balances trajectory accuracy with spectral fidelity, ensuring that temporal coherence is maintained while preventing artificial dissipation. Through the mathematical relationship between frequency-domain accuracy and energy conservation, this approach achieves implicit physics-informed learning—conservation principles emerge naturally from spectral accuracy rather than requiring explicit constraints \citep{karniadakis2021physics}. However, the efficacy of this multi-objective optimization approach is regime-dependent, varying with both system complexity and data availability.

Despite these benefits, an unexpected phenomenon reveals important limitations in the universal applicability of physics-informed additional losses. In Figure \ref{fig:db_broadband_wSpecLoss}, for linear systems (Cases 1–2), PSD shape error exhibits non-monotonic behavior with dataset size when spectrogram loss is employed, achieving minimum error around $2^9$ samples before degrading at $2^{10}$ and $2^{11}$. In contrast, training with MSE loss alone produces monotonic improvement across all dataset sizes. This divergence suggests that multi-objective optimization introduces gradient conflicts that become pronounced at larger dataset scales. For linear dynamical systems, the time-domain and frequency-domain solutions are tightly coupled through the governing differential equations. Therefore, accurate time-domain prediction naturally enforces correct frequency content when sufficient training data are available. The additional spectrogram loss, while beneficial as regularization in data-scarce regimes ($n \leq 2^9$), appears to overconstrain the optimization landscape when the model has sufficient data to implicitly learn the correct spectral characteristics through MSE minimization alone. 

Notably, in Figure \ref{fig:db_broadband_wSpecLoss}, the aforementioned effect is absent in non-linear cases (Cases 3–4), where richer frequency content (harmonics, intermodulation) makes explicit spectral guidance consistently beneficial across all dataset sizes. These results challenge the assumption that physics-informed additional losses universally improve FNO performance, revealing instead a regime-dependent trade-off: what helps guide the model with limited data can interfere with training when sufficient data exist for the model to learn the physics naturally. These results indicate that spectral loss is most beneficial for non-linear systems and small datasets. For linear systems trained on large datasets, it can be reduced or removed since accurate time-domain predictions automatically produce correct frequency content.

In Table \ref{tab:fno_metrics}, the comparison between Broadband and Broadband with Spectrogram Loss reveals interesting observations (Table \ref{tab:fno_metrics}, columns 3 and 4). The inconsistent results exhibit a distribution mismatch between the training and testing sets. The models were trained exclusively on smooth, continuous sinusoidal forcing signals with frequencies between 0.04–2.1\, Hz. The test signals—chirps, impulses, and step changes—contain different frequency characteristics. The FNO has never seen the signals with this frequency content during training and must extrapolate far outside its learned distribution. As the spectrogram loss imposes specific frequency patterns during the training, it worsens the test results errors. During training, the loss function learns to penalize deviations from smooth, harmonic spectral content. When confronted with transients, discontinuities, or sweeping frequencies during testing, the trained models constantly want to reproduce similar patterns. Therefore, the model cannot adapt because it has been explicitly trained to suppress exactly the spectral features these test signals require.

The chaotic performance patterns reflect accidental alignment or misalignment between training and test frequency content (Table \ref{tab:fno_metrics}, columns 3 and 4). When a test signal's frequencies are well represented in the training data, performance is acceptable. When they are different frequencies or require spectral features that the model was learned to suppress, performance collapses. This issue is especially observable in non-linear system test results, where the cubic stiffness term generates harmonics that depend on the input. These different patterns of displacements can produce entirely different mixes of frequencies that the model has never experienced before

\subsection{Discretization Error and Aliasing in Non-linear Systems}







Consider our forcing function from Equation \eqref{eq:force} with two frequencies: $\omega_1$ and $\omega_2$. In a linear system, these frequencies would pass through with modified amplitudes and phases. But the cubic non-linearity $k_3(x_1 - x_2)^3$ generates harmonics which create new frequency content at $n\omega_1 + m\omega_2$ for various integer combinations of $n$ and $m$ which are poorly represented in the FNO. 
Referring to Section \ref{sec:piriori_error}, for our non-linear system, this discretization error becomes a problem. 

Our forcing function $F(t)$ in training is infinitely smooth ($s \approx \infty$) with exponentially decaying Fourier coefficients but the non-linear response at testing may contain many high-frequency components  induced by the harmonics from non-linear springs and/or forcings. 
However, they extend the frequency spectrum far beyond the FNO's retained modes $k_{max}$. As shown by Kovachki et al. \citep{kovachkiUniversalApproximationError2021}, this creates a fundamental lower bound on approximation error proportional to modes $|k| \leq N$.
Furthermore, as \citep{lanthalerDiscretizationErrorFourier2024a} show FNO accuracy degrades as $N^{-s}$ where $s$ is the regularity of the inputs. Our system response to low regularity inputs (e.g. sudden stop and steps) illustrates this loss.

This analysis explains the success of FNO on the IEA 15MW wind turbine dataset. Despite mechanical complexity with hundreds of degrees of freedom, wind turbines exhibit predominantly linear modal behavior. The SCADA acceleration inputs already contain the system's frequency content, and since linear dynamics preserve bandwidth, the FNO can accurately learn the operator mapping without encountering bandwidth expansion or aliasing issues. \citet{kovachkiUniversalApproximationError2021} prove that FNOs approximate linear and quasi-linear PDE operators with polynomial or even sub-linear error scaling, which our wind turbine results validate. The linearity of the offshore wind turbine system keeps the frequency content bounded, enabling the efficient approximation guaranteed by theory.

The spectrogram loss partially mitigates the challenges by addressing the FNO's architectural bias toward low frequencies \citep{oommenIntegratingNeuralOperators2025}. By explicitly reweighting to penalize high-frequency errors, it forces the network to maintain spectral fidelity rather than defaulting to smooth approximations.  However, spectrogram loss cannot overcome the fundamental limitation identified in the theoretical framework -- inadequate representation in the Fourier modes. Information in modes beyond $k_{max}$ is irretrievably lost through truncation. On discrete grids, aliasing further corrupts the retained modes as shown by Lanthaler et al. \citep{lanthalerDiscretizationErrorFourier2024a}. The spectrogram loss removes one error source—spectral bias—but cannot recover truncated information or eliminate aliasing. This is why non-linear systems perform poorly even with frequency-aware training.

\citet{kovachkiUniversalApproximationError2021} shows FNOs can efficiently emulate pseudo-spectral methods for PDEs through clever composition of Fourier and non-linear layers. Here, we showed that when non-linearity causes bandwidth expansion beyond the FNO's representational capacity, accuracy fundamentally degrades regardless of training data quantity or loss function design. This is not a failure of the FNO architecture per se, but more of a fundamental constraint when the operator needs to perform outside the bandwidth it can represent due to Fourier mode truncation.


\FloatBarrier

\section{Conclusion}
\label{sec:conc}
This work set out to chart where Fourier Neural Operators (FNOs) do and do not work for structural dynamics, using experiments that separate the effects of non-linearity, time variation, forcing spectrum, and data scale. We built a controlled testbed around a two-degree-of-freedom (2DOF) mass–spring system with four physics cases (linear; linear with softening; non-linear; non-linear with softening) and three forcing regimes (low, high, broadband), and we trained on nested datasets from $2^6$ to $2^{12}$ samples so scaling trends are cleanly comparable. We complemented these synthetic studies with the IEA 15\, MW reference wind turbine, which is high-dimensional but predominantly linear in the operating regimes considered. Across all of these, the message is unambiguous, which is that FNO performance is governed by \emph{non-linearity}, not system size or dimensional complexity.

On linear dynamics, FNOs behave as claimed. With modest data, they deliver energy ratios near unity and PSD-shape errors below 10\%, and they extend this performance to time-varying stiffness (softening). This is the regime where a truncated Fourier representation captures the relevant physics. The IEA 15\, MW turbine results are a direct confirmation in a realistic, high-DOF setting, where despite hundreds of structural modes and multi-input excitation, the predominantly linear behavior allows the FNO to predict bending-moment time series accurately under both power-production (DLC~1.2) and idling (DLC~6.4) conditions, and to generalize to shutdown (DLC~4.1) even though that transient was not included in training. Energy ratios remain $\approx$1, while spectral coherence lands in the mid-range—consistent with phase sensitivities in rapid transients—yet the amplitude content is captured correctly.

The failure mode emerges as non-linearity injects frequency content beyond the kept modes. In the 2DOF system, the non-linear cubic stiffness generates harmonics and intermodulation products that push energy into higher frequencies. Once that happens, the mode truncation removes exactly the content the model needs to represent, so the network must either ignore it or alias it. 
This produces the experimental features we observed: (i) unchanged energy dissipation in predictions (30–50\% loss for broadband forcing), (ii) PSD-shape errors exceeding 60\% that do not improve with more data, and (iii) a clear plateau beyond $2^9$ training samples, where additional data fail to decrease the error. In short, the error floor is architectural, not statistical.

Frequency content matters as much as the physics. FNOs trained on limited frequency bands behave like unintended frequency-selective filters. When tested out of band, predicted amplitudes collapse—energy ratios can fall to small values, though spectral coherence remains high. That contradiction is important, as the model tracks the \emph{shape} of the frequency distribution while losing \emph{magnitude}. For engineering purposes, this is unacceptable; energy is the quantity that drives stresses, fatigue, and design decisions. Broadband training increases robustness but lowers peak accuracy, which means the model spreads capacity across the spectrum and never becomes truly sharp anywhere. The trade-off is explicit and application-dependent.

We also tested whether training can compensate for architectural bias. A spectrogram-augmented loss, which supervises both magnitude and phase in the time–frequency plane, substantially improves energy preservation (ratios $\approx 1$ in the data-efficient regime) and reduces PSD-shape errors for linear and non-linear cases. That improvement is causal and aligns with theory. It means reweighting FNO weights to penalize frequency-domain error counteracting the low-frequency preference of FNOs and enforces conservation indirectly through Parseval’s relation. However, the benefit is regime dependent. For linear dynamics with larger datasets, the spectrogram term can induce gradient conflict and slightly degrade accuracy relative to plain MSE. When the time- and frequency-domain objectives are already aligned in linear cases, and when data are abundant, so extra constraints can oversteer optimization. For non-linear dynamics, the spectrogram loss helps across data scales but cannot remove the architectural ceiling error set by mode truncation and regularity loss. It reduces the bias; it does not restore information that the representation cannot carry.

The wind turbine shutdown results make the same point from another angle. Training on operation and idling yields models that transfer to shutdown without being shown it explicitly. That only works because the underlying structural response remains predominantly linear. This means the neural network learned the structural operator and can handle a new forcing regime within the same envelope as the function class. The coherence drop during shutdown is consistent with phase fluctuations in short transients and is less important than the accurate energy content. The lesson is not that FNOs ``generalize to everything,'' but that they generalize \emph{within} the space their representation can express.

From these findings, we can be concrete about deployment.

\emph{Use FNOs when:}
(i) the system is linear or weakly non-linear in the operating envelope; 
(ii) the forcing spectrum is band-limited and represented in training; 
(iii) training and deployment distributions are aligned (including transients, if those matter); and 
(iv) fast, resolution-independent inference is a priority. 
In this regime, data needs are modest (hundreds of samples), accuracy is high, and the architectural assumptions match the physics.

\emph{Avoid FNOs when:}
(i) strong non-linearities exist (cubic stiffness, contact, geometric effects) that generate substantial harmonic content; 
(ii) impulses, steps, or sweeping-frequency transients are central to the task; 
(iii) strict energy fidelity is required across a broad spectrum; or 
(iv) training and evaluation spectra differ meaningfully. 
Here, the combination of mode truncation and aliasing sets a hard floor that additional data or minor loss tweaks cannot overcome.

Our research, along with ideas from \citet{kovachkiUniversalApproximationError2021} and \citet{lanthalerDiscretizationErrorFourier2024a}, shows that using FNOs for non-linear systems requires careful analysis of bandwidth. This helps us understand how non-linearities generate different frequencies, and indicates that we need to keep enough modes to ensure the highest frequency, \(k_{max}\), captures all the harmonics created. We must also use a grid resolution that meets the Nyquist criteria for all these frequencies. If bandwidth increases significantly due to the non-linearity, we might need to explore alternative methods, such as multi-scale techniques or hybrid approaches combining FNOs with high-frequency correction terms.


There are different ways this work can be continued. Architecturally, adaptive or learned mode allocation could target frequencies where non-linear energy accumulates. Also, anti-aliasing and de-aliasing strategies can be a standard option. Another approach can be multiscale operators (e.g., Fourier–wavelet hybrids), which can add time–frequency localization to capture local high-frequency content. Hybrid surrogates that use an FNO for the dominant linear map plus a residual network for non-linear corrections are a possible fit for weakly non-linear systems. Additionally, Physics-aware constraints (energy-preserving layers) can mitigate artificial dissipation. On the data side, training should include the transients you care about and broaden amplitude–phase distributions so the model does not overfit to limited spectral patterns. 


In summary, we demonstrated the limitations of FNOs. Non-linearity, rather than complexity, sets the boundary for current FNOs on the class of problems studied here. Recognizing that boundary allows practitioners to choose the right tool and points directly to the architectural and training changes to move the boundary outward.



\section*{Declaration of competing interest}
The authors declare that they have no known competing financial interests or personal relationships that could have appeared to influence the work reported in this paper.

\section*{Acknowledgments}
Funding support for this work from the Tufts Institute for AI and the National Science Foundation, NSF awards GEO2004302 and OAC2137603 is acknowledged. The authors acknowledge the Tufts University High Performance Compute Cluster (https://it.tufts.edu/high-performance-computing) which was utilized for the research reported in this paper.

\bibliographystyle{elsarticle-num-names} 
\bibliography{main}

\end{document}